\documentclass[article,longauth]{aa}

\usepackage{graphicx}
\usepackage{txfonts}

\usepackage[normalem]{ulem}
\usepackage{mhchem}
\usepackage{booktabs}
\usepackage[colorlinks=true, linkcolor=blue, citecolor=blue, urlcolor=blue]{hyperref}

\usepackage{lipsum}
\usepackage{lscape} 
\usepackage{placeins} 
\usepackage{subcaption}

\newcommand{\rev}[1]{\textcolor{black}{#1}}

\begin{document} 

   \title{MINDS: Intertwined evolution of dust and gas in large planet-forming disks}
\subtitle{A diversity driven by halted pebble drift?}
  
  \titlerunning{MINDS: giant T Tauri disks}
  
  \authorrunning{B. Tabone}


   \author{Benoît Tabone\inst{1}
   \and
   Milou Temmink\inst{2}
      \and
   Laurens B. F. M. Waters\inst{3}
   \and
   Ewine F. van Dishoeck\inst{2,4}
         \and
   Andrew Sellek\inst{2} 
            \and
   Pacôme Estève\inst{1} 
         \and
   Nicolas T. Kurtovic\inst{4}
         \and
    Inga Kamp\inst{5}
       \and
    Thomas Henning\inst{6}
        \and
   Danny Gasman\inst{7}
        \and
   Sierra L. Grant\inst{8}
       \and
        József Varga\inst{9,10}
       \and
   Alice Guerras\inst{1}
        \and
   Dmitry Semenov\inst{11,6}
   \and
    Aditya M. Arabhavi\inst{5}
        \and
    Alessio Caratti o Garatti\inst{12}
    \and
    Anne Dutrey\inst{13}
    \and
    Edwige Chapillon\inst{13,14}
    \and
    Stéphane Guilloteau\inst{13}
    \and
    Manuel Güdel\inst{15,16}
     \and
    Hyerin Jang\inst{3}
         \and
    Till Kaeufer\inst{17}
    \and
    Jayatee Kanwar\inst{5,18,19,23}
     \and
    Göran Olofsson\inst{20}
    \and
    Giulia Perotti\inst{21}
    \and
    Vincent Piétu\inst{14}   
       \and
    Thomas P. Ray\inst{22}
    \and
    Marissa Vlasblom\inst{2}
          }


            \institute{
   Université Paris-Saclay, CNRS, Institut d’Astrophysique Spatiale, 91405 Orsay, France
   \and
   Leiden Observatory, Leiden University, PO Box 9513, 2300 RA Leiden, The Netherlands
   \and
Department of Astrophysics/IMAPP, Radboud University, PO Box 9010, 6500 GL Nijmegen, The Netherlands
   \and
    Max-Planck-Institut für Extraterrestrische Physik, Giessenbachstrasse1, 85748 Garching, Germany 
    \and
Kapteyn Astronomical Institute, Rijksuniversiteit Groningen, Postbus 800, 9700AV Groningen, The Netherlands
    \and
    Max-Planck-Institut für Astronomie (MPIA), Königstuhl 17, 69117, Heidelberg, Germany
    \and
Institute of Astronomy, KU Leuven, Celestijnenlaan 200D, 3001 Leuven, Belgium
\and
Earth and Planets Laboratory, Carnegie Institution for Science, 5241 Broad Branch Road, NW, Washington, DC 20015, USA
    \and
HUN-REN Research Centre for Astronomy and Earth Sciences, Konkoly Observatory,
MTA Centre of Excellence, Konkoly Thege Miklós út 15-17., H-1121 Budapest, Hungary
    \and
    CSFK, MTA Centre of Excellence, Konkoly-Thege Miklós út 15-17, H-1121 Budapest, Hungary
    \and
Zentrum f\"{u}r Astronomie der Universit\"{a}t Heidelberg, Institut f\"{u}r Theoretische Astrophysik, Albert-Ueberle-Str. 2, 69120 Heidelberg, Germany
\and
INAF – Osservatorio Astronomico di Capodimonte, Salita
Moiariello 16, 80131 Napoli, Italy
\and
Laboratoire d'Astrophysique de Bordeaux, Université de Bordeaux, CNRS, B18N, Allée Geoffroy Saint-Hilaire, F-33615 Pessac, France
\and
Institut de Radioastronomie Millimétrique (IRAM), 300 rue de la Piscine, F-38406 Saint-Martin d'Héres, France
\and
Dept. of Astrophysics, University of Vienna, Türkenschanzstr. 17,
A-1180 Vienna, Austria
\and
ETH Zürich, Institute for Particle Physics and Astrophysics,
Wolfgang-Pauli-Str. 27, 8093 Zürich, Switzerland
\and
Department of Physics and Astronomy, University of Exeter, Exeter EX4 4QL, UK
\and
Space Research Institute, Austrian Academy of Sciences, Schmiedlstr. 6, 8042, Graz, Austria
\and
 Institute for Theoretical Physics and Computational Physics, Graz University of Technology, Petersgasse 16, 8010, Graz, Austria
 \and
Department of Astronomy, Stockholm University, AlbaNova University Center, 10691, Stockholm, Sweden
\and
Niels Bohr Institute, University of Copenhagen, NBB BA2, Jagtvej 155A, 2200 Copenhagen, Denmark
\and
Dublin Institute for Advanced Studies, 31 Fitzwilliam Place, D02 XF86 Dublin, Ireland
\and
Department of Astronomy, University of Michigan, 1085 S. University, Ann Arbor, MI 48109, USA
            }     
    
\date{\today}
\abstract
{JWST gives unique access to the chemical and physical conditions in inner disks ($< 10~$au) where the majority of detected exoplanets are thought to form.
}
{Our goal is to investigate the diversity of inner disks with a specific focus on large and massive disks around Sun-like stars. These are thought to be the progenitors of planetary systems with wide-orbit planets and potential instances of disks with quenched pebble drift. 
}
{We analyze the MIRI-MRS spectra of three disks among the MINDS program orbiting young Sun-like stars ($\simeq 0.8~M_{\odot}$): V1094 Sco, DL Tau, and IM Lup, which are put in context with the MINDS sample and millimeter observations.
}
{The JWST spectra reveal a striking diversity, in line with previous observations. V1094 Sco and DL Tau exhibit the highest C$_2$H$_2$/H$_2$O flux ratio of the MINDS sample of full T Tauri disks ($M_* > 0.4 M_{\odot}$). In V1094 Sco, even cold C$_4$H$_2$ is seen. In contrast, the IM Lup spectrum is dominated by O-bearing species. No one-to-one correspondence is found between the gas in the outer disk, as traced by the C$_2$H (3-2)/C$^{18}$O (2-1) flux ratio, and that of the inner disk as traced by the C$_2$H$_2$/H$_2$O flux ratios. To account for these observational results, we propose a scenario based on a toy model of halted pebble drift relevant to the pebble-rich disks. We show that a volatile C/O ratio close to unity and low elemental C and O abundance can be achieved in inner disks only if: (1) about 95$\%$ of the icy grains are blocked in the outer disk, (2) the outer disk is chemically evolved to reach elevated gas-phase C/O and low (C+O)/H abundance ratio, and (3) the gas in the outer disk had time to be advected to the inner disk. In this scenario, DL Tau and perhaps V1094 Sco would be the rare examples for which all these conditions are met. Therefore, a high C$_2$H$_2$/H$_2$O flux ratio in pebble-rich disks would have a different origin than proposed for very-low mass stars, for which fast drift of oxygen-rich pebbles would eventually leave a carbon-rich inner disk. Destruction of carbonaceous grains appears to be a less compelling scenario to account for the high C$_2$H$_2$/H$_2$O flux ratio in DL Tau and V1094 Sco because existing models require strong pebble drift and low accretion rates to enrich the inner disk in volatile carbon. 
Alternatively, the mid-IR molecular features could be related to low-optical depth of the dust, which would reveal a normally hidden reservoir of C$_2$H$_2$ deeper down to the midplane.
Finally, we show for the first time that the disks with \rev{high C$_2$H$_2$/H$_2$O flux ratio} exhibit a prominent silica dust component, a result found in four disks published so far (V1094 Sco, DL Tau, CY Tau, DoAr 33). We propose that the reformation of dust at the sublimation front of silicates in a gas with super-solar (but below unity) C/O ratio leads to a silica stoichiometry (SiO$_2$). In turn, silica constitutes a promising diagnostic of the C/O ratio in the inner disks.}
{The diversity of inner disks as revealed by mid-IR spectroscopy could be driven by the joint evolution of the outer versus inner disk via chemical evolution of the cold outer disk and radial redistribution of gas and ice-coated dust. Studies gathering large samples and forward modelling are required to confirm this proposal and obtain a global picture of disk evolution.}           
\keywords{Protoplanetary disks -- Astrochemistry -- Infrared: planetary systems -- Astrochemistry -- Radiative transfer}

\maketitle


\section{Introduction}

Exoplanets are common, testifying that planet formation is a widespread phenomenon in disks orbiting young stars. The most striking results obtained by the exoplanet community are the large diversity in the properties of exoplanets and the occurrence of specific kinds of planetary systems \citep{2019ApJ...874...81F,2023ASPC..534..839L}. The next challenge is to understand how different evolutionary pathways of planet-forming disks lead to the diversity and occurrence of planetary systems \citep[][]{2016JGRE..121.1962M,2023ASPC..534..717D}. In the next decade, constraints on the elemental composition
of gaseous exoplanets will be gathered on statistically representative samples thanks to ambitious space missions like JWST \citep[e.g.,][]{2025arXiv250601800W,2025AJ....170...61M} and ARIEL \citep{2018ExA....46..135T}, and ground-based observations \citep[e.g., VLTI-GRAVITY, VLT-CRIRES+, ELT-METIS,][]{2025arXiv250508926S}. This opens a new avenue for understanding the formation history of some exoplanetary systems \citep{2016ApJ...832...41M,2019ARA&A..57..617M,2022ApJ...934...74M}. 

In disks, the elemental composition of the gas and solids is indeed expected to vary in space and time due to various processes such as condensation of volatile species, radial and vertical mixing, radial drift of icy grains or destruction of carbonaceous grains \citep{2011ApJ...743L..16O,2017MNRAS.469.3994B,2020ApJ...899..134K}. This implies that the elemental composition of exoplanets carries information about where and when planets form. Several sophisticated models have been explored to predict the distribution of elements during the disk lifetime and the resulting composition of exoplanets \citep[e.g.,][]{2019A&A...627A.127C}. Yet, considering the complexity of the processes setting the disk properties, direct observations are critically needed to determine the actual composition of disks and enable the interpretation of exoplanet's compositions \citep{2022ApJ...934...74M,2024ApJ...969L..21B}.

The \textit{Herschel} Space Observatory, along with ground-based millimeter facilities such as ALMA and NOEMA, has enabled detailed studies of the gas composition in the cold outer regions of disks ($>~20~$au). 
Weak H$_2$O and CO lines point toward low abundances of gas-phase CO and water \citep{2010A&A...521L..33B,2013ApJ...776L..38F,2017A&A...599A.113M}. This was interpreted as the result of a combination of freeze-out of volatiles onto dust grains, followed by dust growth and drift \citep{2017ApJ...842...98D,
2018ApJ...864...78K}, and chemical conversion of CO into less volatile species such as CH$_3$OH, CO$_2$, or CH$_4$ \citep{2018A&A...618A.182B,2020ApJ...899..134K,2022ApJ...938...29F}. The sequestration of volatile C and O would occur on a relatively short timescale of the order of 1~Myr \citep{2020ApJ...891L..17Z} and is accompanied by an increase in the C/O ratio in the gas as indicated by the \rev{millimeter} emission of small hydrocarbons \citep{2016ApJ...831..101B}.

However, the majority of characterized exoplanets form in the inner 10~au, a region best probed by mid-IR wavelengths. Space-borne and ground-based instruments operating in the infrared have demonstrated the power of IR spectroscopy to unveil the chemical composition of inner disks with the detection of the main carbon and oxygen carriers like H$_2$O, CO, CO$_2$, OH, or HCN in warm gas \citep[$\gtrsim 300~$K][]{2008Sci...319.1504C,2010ApJ...720..887P,
2011ApJ...731..130S}. Today JWST brings the planet formation domain closer to the exoplanet domain by providing unique spectroscopic capabilities from 1 to 27 $\mu$m \citep{2024PASP..136e4302H,2024ApJ...963..158P}. The exquisite sensitivity opens the possibility of characterizing faint disks \citep[e.g.,][]{2023NatAs...7..805T,2024Sci...384.1086A,2025ApJ...978L..30L} and the good spectral resolution enables a detailed study of the excitation of the species and their isotopologues \citep[e.g.,][]{2023ApJ...947L...6G, 2023A&A...679A.117G,2024ApJ...975...78R,2024A&A...686A.117T,2025AJ....169..184S, 2025arXiv250713921F}. 

Considering the richness of the mid-IR spectral region, the first studies of JWST-MIRI spectra were conducted on individual sources, highlighting the diversity in the mid-IR emission, depending on the physical structure of their disks \citep[e.g.,][]{2023Natur.620..516P} and the stellar mass \citep[e.g.,][]{2023NatAs...7..805T}. The recent studies based on larger samples quantify this diversity focusing on fluxes of specific molecular features. \rev{Following first trend found with \textit{Spitzer} \citep[][]{2017ApJ...834..152B}, disks with cavities 
appear to be overall distinct in molecular emission \citep[][]{2026ApJ...998..255M}, underpining the role of the physical conditions in inner disks.} Regarding the stellar mass, \citet{2025arXiv250602748A,2025arXiv250804692G}, show a steep decline in the C$_2$H$_2$/H$_2$O flux with increasing stellar mass, a trend hinted at by \textit{Spitzer} observations \citep{2013ApJ...779..178P}.

However, the stellar mass or \rev{cavities are not the only parameters setting the emission of inner disks}. In a companion paper, \citet{2025arXiv250804692G} revealed that for a given stellar mass, the C$_2$H$_2$ over H$_2$O flux ratio varies by more than an order of magnitude, \rev{a result also found by the JDISCS collaboration \citep{2025arXiv250507562A}}. A compelling scenario to set the chemical composition of inner disks is the radial transport of chemical elements due to the differential transport of semi-volatile species (H$_2$O, CO$_2$, CH$_4$) as ice through dust radial drift \citep{2017MNRAS.469.3994B,2023A&A...677L...7M,2025A&A...694A..79S}. If icy pebbles were to be trapped in some disks and at different locations, as hinted at based on \textit{Spitzer}-IRS observations \citep{2013ApJ...766..134N}, this could drive the observed diversity \citep{2024A&A...686L..17M,2021ApJ...921...84K}. Variation in the cosmic ray ionisation rate, which controls the chemical conversion of CO, \rev{could also be a} key driver of this diversity \citep{2025arXiv250711631S}. Indirect \rev{signatures} of the drift of icy pebbles are seen in the form of cold water ($\simeq 200~$K) \rev{which is interpreted as the result of sublimation of drifting icy-pebbles followed by gas diffusion toward the disk atmosphere} \citep{2023ApJ...957L..22B,2024ApJ...975...78R,2025ApJ...990L..72K,2025arXiv250515237T}.
Expanding to HCN, H$_2$O, C$_2$H$_2$, \citet{2025A&A...694A.147G} showed that relating the chemical composition of the inner disk to the presence of dust traps is complicated by the limited spatial resolution of ALMA to map the sublimation regions of the main O- and C-carriers. \rev{More recent analysis suggest that pebble delivery could be regulated by the inner-most dust traps \citep{2025ApJ...990L..72K}.} The destruction of carbon grains, which would enrich the gas-phase in carbon, could also drive this diversity, depending on the outward transport of gas-phase carbon  \citep{2024ApJ...977..173C,2025A&A...699A.227H}.

In this study, we aim to characterize and discuss the diversity inner disks for which pebble-drift is thought to be halted or at least slowed down. To achieve this goal, we study the MIRI-MRS spectra of three disks among the MINDS program \citep{2024PASP..136e4302H}: IM Lup, DL Tau, and V1094 Sco. These disks are orbiting stars of similar mass ($\simeq 0.7-0.9~M_{\odot}$) and are among the 3$\%$ largest disks in terms of millimeter gas and dust emission \citep[e.g.,][]{2018A&A...616A..88V}, \rev{and do not show any sign of a cavity}. The source of diversity is therefore minimized, whereas these disks are the best candidates to retain icy-grains in their outer regions. Moreover, targeting large disks allows us to strengthen the connection with sub-samples of planetary systems as these large disks are likely the progenitors of systems with wide-orbit planets \rev{similar to} HR 8799 \citep{2008Sci...322.1348M} or HD 95086 \citep{2013ApJ...772L..15R}. \rev{The disks studied here are put in context with the MINDS sample, selecting full disks orbiting T Tauri stars, for which C$_2$H$_2$ and H$_2$O fluxes have already been measured by \citet{2025arXiv250804692G}. 
}
In Sec. \ref{sec:observations}, we present the data reduction and the slab model used to analyze the data, and in Sec. \ref{sec:results} we analyze the molecular and dust features. The results are discussed in light of the molecular emission of the outer disks \rev{using an expanded sample of pebble-rich disks} and of a toy model applicable to halted pebble-drift in Sec. \ref{sec:discussion}. In this section, we also report on the discovery of a connection between the composition of the silicate dust and the gas composition. Our findings are summarized in Sec. \ref{sec:conclusions}.

\section{Observations and analysis method}
\label{sec:observations}

\begin{table*}
\begin{center}
 \caption{Properties of the host stars and the disk for the selected sample of giant disks.}
\begin{tabular}{ |c|ccccccccc| } 
 \hline
        &  $T_{\rm{eff}}$ & $L_*$ $^{(a)}$ & $\text{log}(L_{\text{acc}})$ $^{(a)}$ & $M_*$ $^{(a)}$& $\log{\dot{M}_*}$ $^{(a)}$ & $R_{\text{CO}}$ & $M_{\text{dust}}$  $^{(a)}$ & $R_{\text{dust}}$ & $\alpha_{13-26}$\\ 
       & [K] & [$L_{\odot}$] & [$L_{\odot}$] & [$M_{\odot}$] & [$M_{\odot}/yr$] & [$\rm{au}$] & [$M_{\oplus}$] & [$\text{au}$] & \\
        \hline
 DL Tau       & 4162 & 1.5 & -0.3 & 0.77 & -7.2 & 597 $^{(d)}$     & 123 & 144 $^{(e)}$ & -0.59 \\ 
 IM Lup / Sz82 & 4210 & 2.5 & -1.1 & 0.87 & -7.9 & 803 $^{(d)}$     & 125 & 244 $^{(e)}$ & -0.06 \\ 
 V1094 Sco     & 4115 & 1.2 & -1.0 & 0.73 & -7.9 & 997 $^{(c)}$   & 135 & 381 $^{(c)}$ & -0.58 \\ 
 \hline
\end{tabular}
\label{table:source_properties}
\end{center}
{(a) \citet{2023ASPC..534..539M} (b) \citet{2021A&A...649A..19S} (c) \citet{2025arXiv250610734D} (d) \citet{2022ApJ...931....6L} (e) \citet{2025arXiv250610746V} (f) \citet{2024A&A...687A.174F}}
\end{table*}

\begin{figure*}[!t]
\centering
\includegraphics[width=0.9\textwidth]{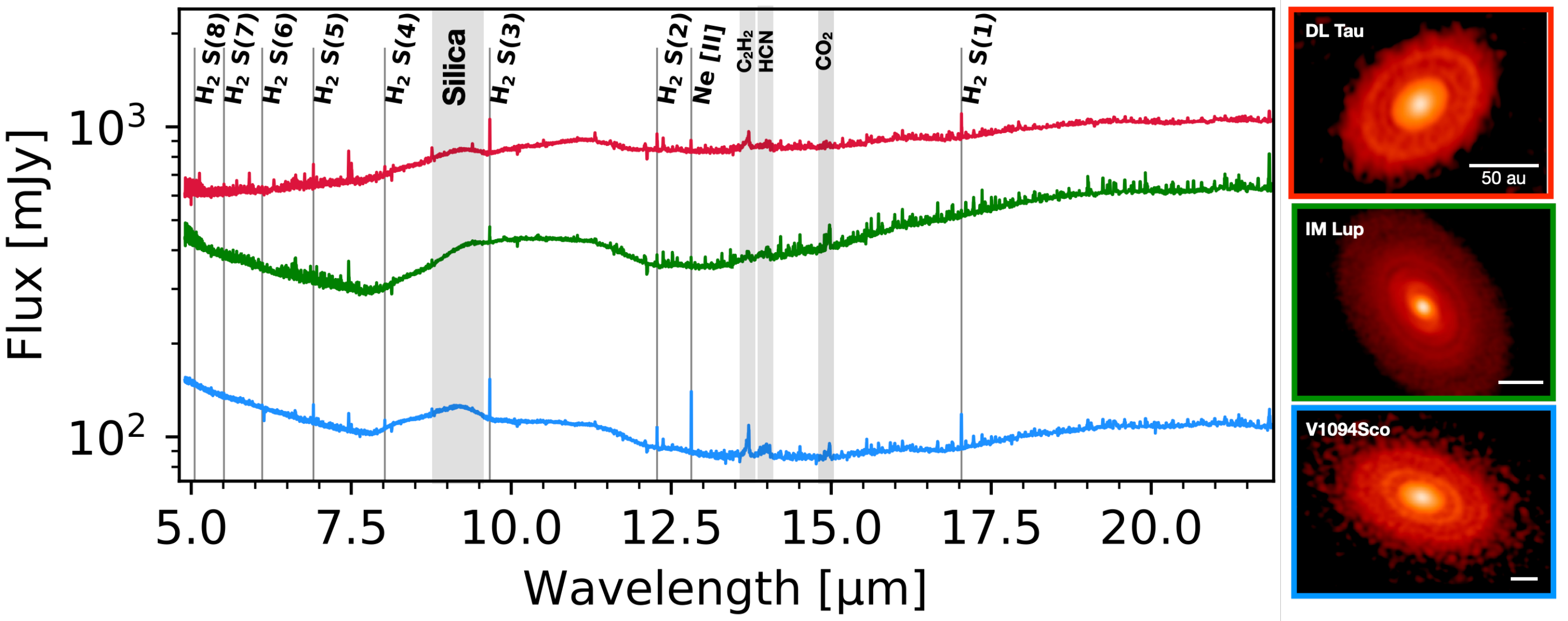}
\caption{Overview of the MIRI-MRS spectra of the three T Tauri disks along with ALMA views of their dust emission for illustrative purposes. The mid-IR emission is dominated by a dust continuum with particularly flat silicate features in DL Tau and V1094 Sco, revealing large micron-sized grains with a prominent silica feature. The 13-16~$\mu$m region shows a wealth of molecular features with a prominent C$_2$H$_2$, HCN, and CO$_2$ features outlined in Fig. \ref{fig:overview-zoom-15mic}. In addition the pure rotational lines of H$_2$ and the Ne [II] line are seen in all three disks.}
\label{fig:overview}
\end{figure*}

\subsection{Data reduction}
Similar to \citet{2025A&A...699A.134T}, the MIRI-MRS observations have been reduced using a standard pipeline \citep[version 1.16.1;][]{BushouseEA24} and pmap 1315. The observations were taken in the FASTR1 readout mode and made use of the four-point dither pattern, using all three grating settings (A, B, and C). From the reduced spectral cubes, the spectra were extracted using standard pipeline methods, where the spectra have been extracted using an aperture equal to two times the full width at half maximum (FWHM) and backgrounds were subtracted using annuli. Furthermore, we used the residual fringe correction option of the standard pipeline, as two of our three sources (DL Tau and IM Lup) were taken without target acquisition. We further subtract the dust and gas continuum emission following the method of \citet{2024A&A...686A.117T} who used Savitzky-Golay filters as implemented by \citet{2022zndo...7255880E}. 

\subsection{Slab model fit}

The molecular features are analyzed using the single-zone LTE slab model presented in \citet{2023NatAs...7..805T} and widely used within the MINDS collaboration \citep[e.g.,][]{2023Natur.620..516P,2023ApJ...947L...6G,2023A&A...679A.117G,2025arXiv250515237T}. The energy levels are assumed to be populated according to a Boltzmann distribution with a temperature $T$ and the intrinsic line profile function to be Gaussian with a broadening of $\sigma = 2~$km~s$^{-1}$ (full-width half maximum of $4.7~$km~s$^{-1}$). The calculation of the emerging line intensity involves mutual line overlap by summing the wavelength-dependent optical depth of each line before calculating the specific intensity. This is particularly crucial for $Q$-branches where many lines can screen themselves. However, we neglect the mutual line overlap between different species except for C$_2$H$_2$ and its isotopologue $^{13}$CCH$_2$ which is thought to share the same emitting region and have overlapping lines. The list of levels and radiative transitions is extracted from the HITRAN database \citep{2022JQSRT.27707949G}.

The three free parameters of a single-slab model are the excitation temperature $T$, the column density of the species $N$, and the effective emitting radius $R$ which sets the emitting area $A$ with the convention $A= \pi R^2$. We recall that the flux is directly proportional to $R^2$ but also, in the optically thin regime, to $N$. Following the calculations of the emergent flux (typically measured in Jansky), the synthetic spectra are convolved by the spectral response of the MRS and resampled on the same wavelength grid as the observed spectrum using the \texttt{spectres} library \citep{SpectRes}. 
The analysis of JWST spectra is often hampered by a pseudo-continuum due to highly optically thick gas ; this cannot be unambiguously separated from the dust continuum \citep[see e.g., J160532 and ISO-ChaI-147,][]{2023NatAs...7..805T,2024Sci...384.1086A}. If one assumes that the continuum exclusively arises from dust, the column densities of molecules can be severely underestimated. \citet{2024A&A...687A.209K} proposed to fit simultaneously the dust and gas continuum but this approach requires fitting the dust emission at the percent level when the line-to-continuum ratio is low, which remains difficult to achieve. In this paper, we propose a new approach. The slab models of the gas are treated as any observed JWST spectrum by subtracting the continuum as in \citet{2024A&A...686A.117T}. This means that the parameters are estimated only based on the amplitude of the molecular features and not on a subjective determination of the dust continuum emission or on a dust model. \rev{The fits of the JWST spectra are performed using an MCMC approach with the affine-invariant ensemble sampler implemented in the \texttt{emcee} package \citep{2013PASP..125..306F} from a grid of precalculated slab models, which are interpolated in $N$ and $T$ during the parameter exploration (see Appendix \ref{app:slab_fit}).} 
In the grids, the temperature and $\log(N)$ are sampled with steps of 25 K and 0.2, respectively. For C$_2$H$_2$ and H$_2$O, a single slab model gives a poor fit and we fit the features with a two or three slabs by summing the flux of each slab. The features are fitted sequentially as in \citet{2023ApJ...947L...6G} starting with H$_2$O, then CO$_2$, OH, C$_4$H$_2$, HCN, and C$_2$H$_2$.


\section{Results}
\label{sec:results}

\begin{figure}
\centering
\includegraphics[width=0.4\textwidth]{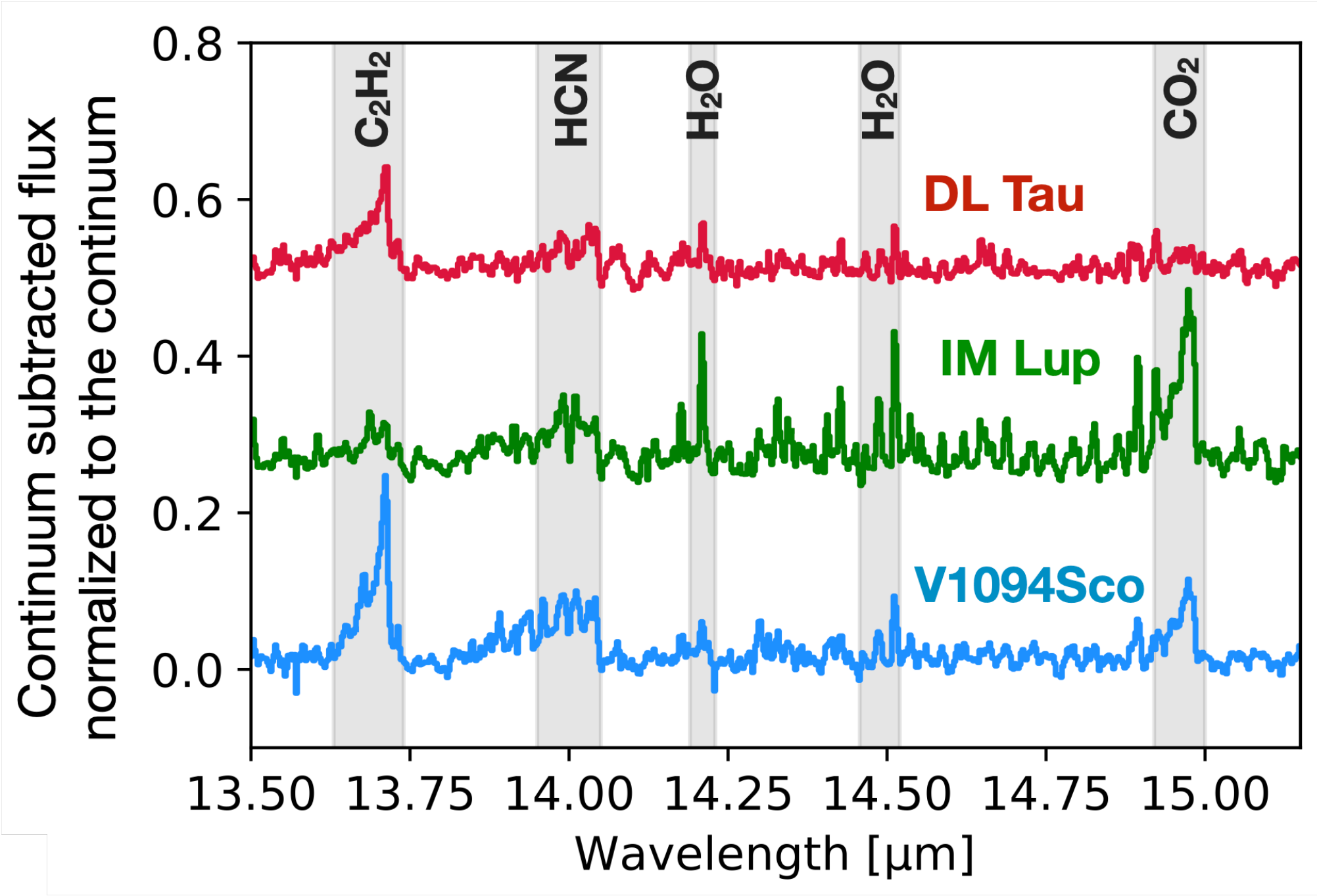}
\caption{Zoom in on the 13.5-15$~\mu$m region showing the main detected species. The continuum subtracted spectra are normalised to the continuum flux at $13~\mu$m and shifted by 0.25 and 0.5 for IM Lup and DL Tau, respectively. A broad diversity in strength of organic features is found between the three disks. The shaded areas are only indicative of regions where the molecular features of the species are particularly bright.}
\label{fig:overview-zoom-15mic}
\end{figure}

\begin{figure}
\centering
\includegraphics[width=0.46\textwidth]{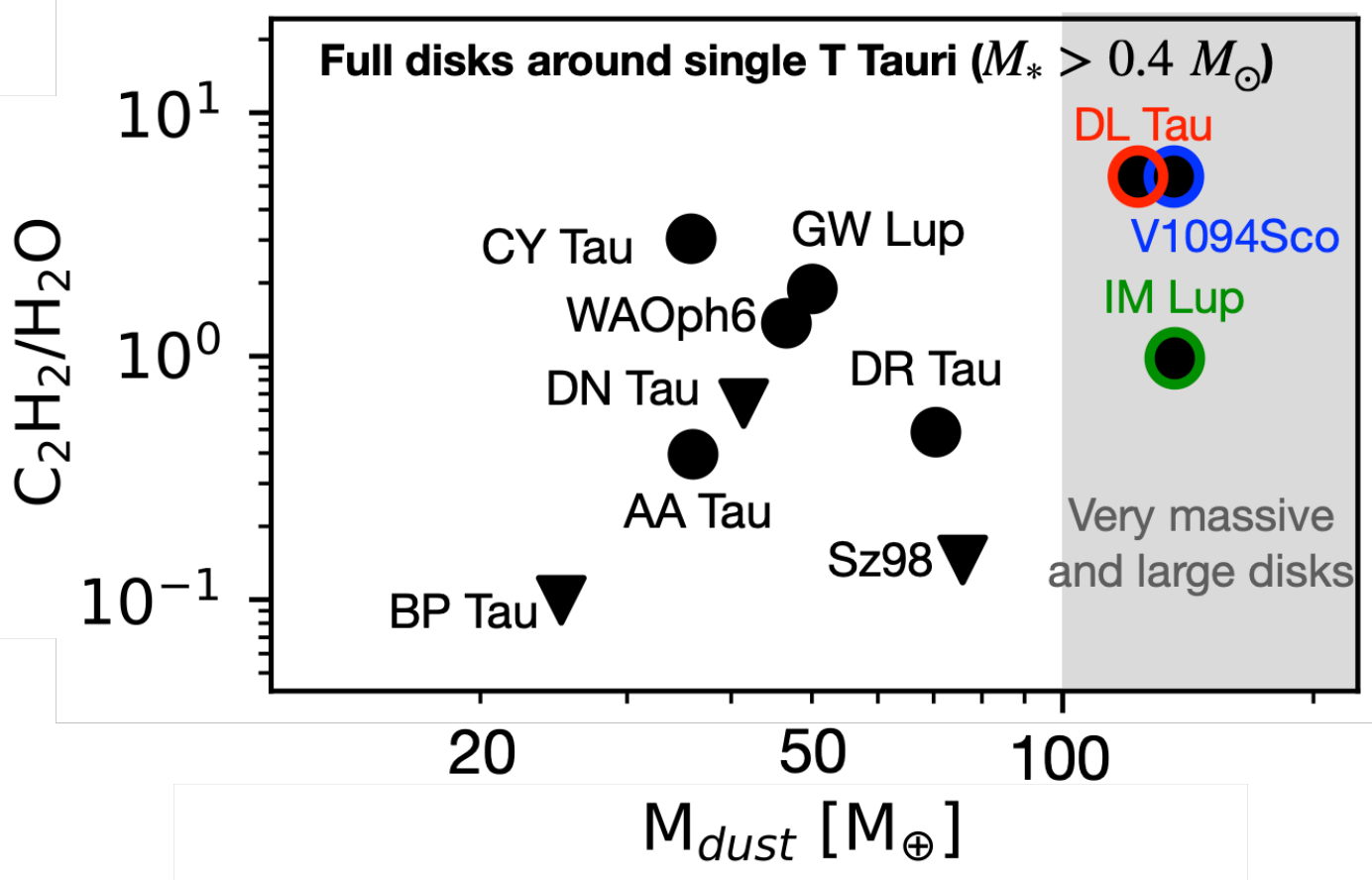}
\caption{C$_2$H$_2$/H$_2$O flux ratio versus dust mass of the MINDS sample of full disks orbiting single T Tauri stars \citep{2025arXiv250804692G}.
Our study is focused on three giant disks that contain \rev{the largest amounts of dust in their outer regions among the MINDS sample}. DL Tau and V1094 Sco are the brightest disks in C$_2$H$_2$ with respect to H$_2$O within the sample. 
}
\label{fig:context}
\end{figure}

\subsection{Overview of the MIRI-MRS spectra}

\label{subsec:overview}

\subsubsection{The sample}

The properties of the three sources analyzed in this work are summarized in Table \ref{table:source_properties}.  To reveal \rev{the intrinsic diversity} in disk emission and not be driven by correlations with stellar mass and outer disk properties, we selected sources with very similar effective temperatures and outer disk properties. The selected spectral type is K7-6, which corresponds to a stellar mass of about 0.7-0.9~$M_{\odot}$. \rev{The three sources also exhibit comparable accretion luminosities, indicating that the differences observed in the mid-IR are not driven by the shallow correlation between $L_{acc}$ and molecular emission \citep{2017ApJ...834..152B,2020ApJ...903..124B}.} Regarding the properties of the outer disks, the dust disk mass inferred from the mm continuum emission is about 120-130~$M_\oplus$, corresponding to a total gas mass of $4 \times 10^{-2}~M_{\odot}$ assuming a gas-to-dust mass ratio of 100. Using the Lupus star-forming region as a reference, our sources are within the top 3$\%$ most massive and largest disks \citep{2018A&A...616A..88V}. 
These disks also belong to the same sub-population of giant disks in the gas with CO gas sizes, as defined by the radius encompassing 90$\%$ of the CO mm flux, larger than 600~au. Using the archival work of \citet{2022ApJ...931....6L}, which is biased toward large disks, this corresponds to the top 10$\%$ of the largest disks in CO millimeter emission.
 
\subsubsection{Continuum emission}

The MIRI-MRS spectra of the three sources are presented in Fig. \ref{fig:overview}. The mid-IR continua span one order of magnitude in absolute flux with V1094 Sco having the weakest emission at the 100~mJy level around 13~$\mu$m. This large variation is not due to the host star since they have similar bolometric luminosity but should originate from the structure of the inner disk. The fluxes of IM Lup and DL Tau are well within the typical mid-IR flux measured in full T Tauri disks orbiting stars of similar SpT \citep{2011ApJS..195....3F}. The mid-IR continuum of V1094 Sco is much dimmer \citep{2011ApJ...726...45T}, pointing toward a flattened disk geometry \rev{limiting the illumination of the dust by the star and from the dust inner rim} \citep{2004A&A...417..159D,2005A&A...434..971D}.

The silicate features of the MINDS sample of T Tauri disks will be analyzed in detail in a forthcoming study (Varga et al. submitted). We note that DL Tau and V1094 Sco have similar 10~$\mu$m silicate features characterized by a weak and flat profile \rev{\citep[see Fig. 3 in][]{2025arXiv250804692G}}. This points to large and well-settled grains \citep{2009ApJS..182..477S,2010A&A...520A..39O}. \rev{This is also supported by the relatively flat SEDs, with 13-26~$\mu$m spectral indices that lie among the lowest 25$\%$ compared to, e.g., samples from Taurus \citep[see Fig. 12 in][]{2006ApJS..165..568F}}. In addition, a peak around 9~$\mu$m attributed to amorphous silica grains (glass of SiO$_2$ stoichiometry) is observed in both sources. Silica-rich grains were identified by \citet{2009ApJ...690.1193S} based on \textit{Spitzer} data. In particular, these authors measured a fraction of warm silica of 25\% in DL Tau, one of the highest fractions in their sample of 65 T Tauri disks. In Section \ref{sec:discussion}, we relate this specific stoichiometry to the gas composition and propose that this \rev{stoichiometry} is due to the formation of silica in gas with elevated C/O elemental ratio. In contrast, IM Lup shows a slightly more peaked silicate feature with a rising SED from 14 and 22~$\mu$m consistent with finer grains and a flared disk geometry.

\subsubsection{Molecular lines}

Molecular emission from organics and water is shown in Fig. \ref{fig:overview}, though with a relatively low contrast between the lines and the continuum compared to emblematic sources like DR Tau \citep{2024A&A...689A.330T} but not at odds with the MINDS sample of T Tauri disks \citep{2025arXiv250804692G}. Figure \ref{fig:overview-zoom-15mic} shows a zoom-in on the 13.5-15.1~$\mu$m region, which is the richest in molecular features. The spectra being normalized by the continuum flux at 13~$\mu$m, it is apparent that the overall line-to-continuum ratios are comparable between the three sources despite the order-of-magnitude differences in the absolute line flux and continuum strength. This suggests that V1094 Sco is a scaled-down version of the other disks, \rev{where the mid-IR emission is simply weaker, likely due to disk geometry which limits the irradiation of the disk}. Yet, when comparing specific features, the three spectra are very different with IM Lup showing prominent emission of O-bearing species, notably CO$_2$, whereas V1094 Sco and DL Tau are dominated by emission from C-bearing species, notably C$_2$H$_2$.

\rev{Figure \ref{fig:context} puts the three giant disks in the context of the MINDS sample analyzed by \citet{2025arXiv250804692G}, excluding the M-dwarfs, the transition disks, and the binary sources,} and focusing on the C$_2$H$_2$ and H$_2$O line fluxes. \rev{Details on the measurement of the C$_2$H$_2$ and H$_2$O flux can be found in \citet{2025arXiv250804692G} where H$_2$O model flux is removed before C$_2$H$_2$ flux is measured. The H$_2$O line flux includes lines with upper energy levels ranging from 2400 to 6000 K and typically traces gas around 400~K. We selected C$_2$H$_2$ because it is the only organic carefully measured over the full MINDS sample and is also proposed to be one of the most compelling species to trace variation in the elemental abundance ratio (see discussion Sec. \ref{subsec:elemental_abundance}). }

\rev{From Fig. \ref{fig:context}, it apparent that the disks of DL Tau and V1094 Sco are the brightest in C$_2$H$_2$ among the MINDS sample of full T Tauri disks when normalized to H$_2$O. Those two sources suggest a correlation between the dust mass and the C$_2$H$_2$/H$_2$O flux ratio when focusing on full disks with stellar mass above $0.4~M_{\odot}$ but Fig. \ref{fig:context} also shows that the size of our curated sample is too small to provide statistically significant results. If present, such a correlation would be reminiscent of the HCN/H$_2$O versus dust mass trend uncovered by \citet{2013ApJ...766..134N}. }

\citet{2020ApJ...903..124B} discussed correlations using the disk size instead of dust mass, showing that within the \textit{Spitzer} sample with existing mm-data, the correlation is dominated by the negative correlation between H$_2$O line fluxes and dust size \rev{with a spread of about one dex}. \rev{With the jump in sensitivity and spectral resolution, JWST converts \textit{Spitzer}-IRS upper limits into detections, and these correlations are expected to be revised in future studies disentangling the effect of outer disk properties, inner cavities, stellar mass, and multiplicity.} 
This will require extensive studies combining different programmes with consistent spectral analysis, a work that it beyond the scope of the present paper. 

\rev{Here, we indeed focus on the largest disks of the MINDS sample and we shall simply conclude that of the three giant disks, two are outliers in terms of C$_2$H$_2$/H$_2$O hinting at a higher occurence rate of elevated C$_2$H$_2$/H$_2$O flux ratio in pebble-rich disks.}
Even if present, a putative correlation will also be associated with a large spread and IM Lup shows a lower C$_2$H$_2$/H$_2$O line ratio, more in line with the bulk sample. Sz 98, a large disk studied by \citet{2023A&A...679A.117G}, represents the other extreme in this spread with very low C$_2$H$_2$/H$_2$O. \rev {Such a diversity is consistently found in surveys like in the recent study of \citet{2025arXiv250507562A}. Figure \ref{fig:context} further illustrates that this diversity remains even when selecting stars and disks that are similar.}

In the following, we study the properties of the molecular emission of the three giant disks using slab models. The best fit parameters are provided in Tables \ref{table:best_fit_slab} and \ref{table:best_fit_slab2}. 

\begin{figure}
\centering
\includegraphics[width=0.5\textwidth]{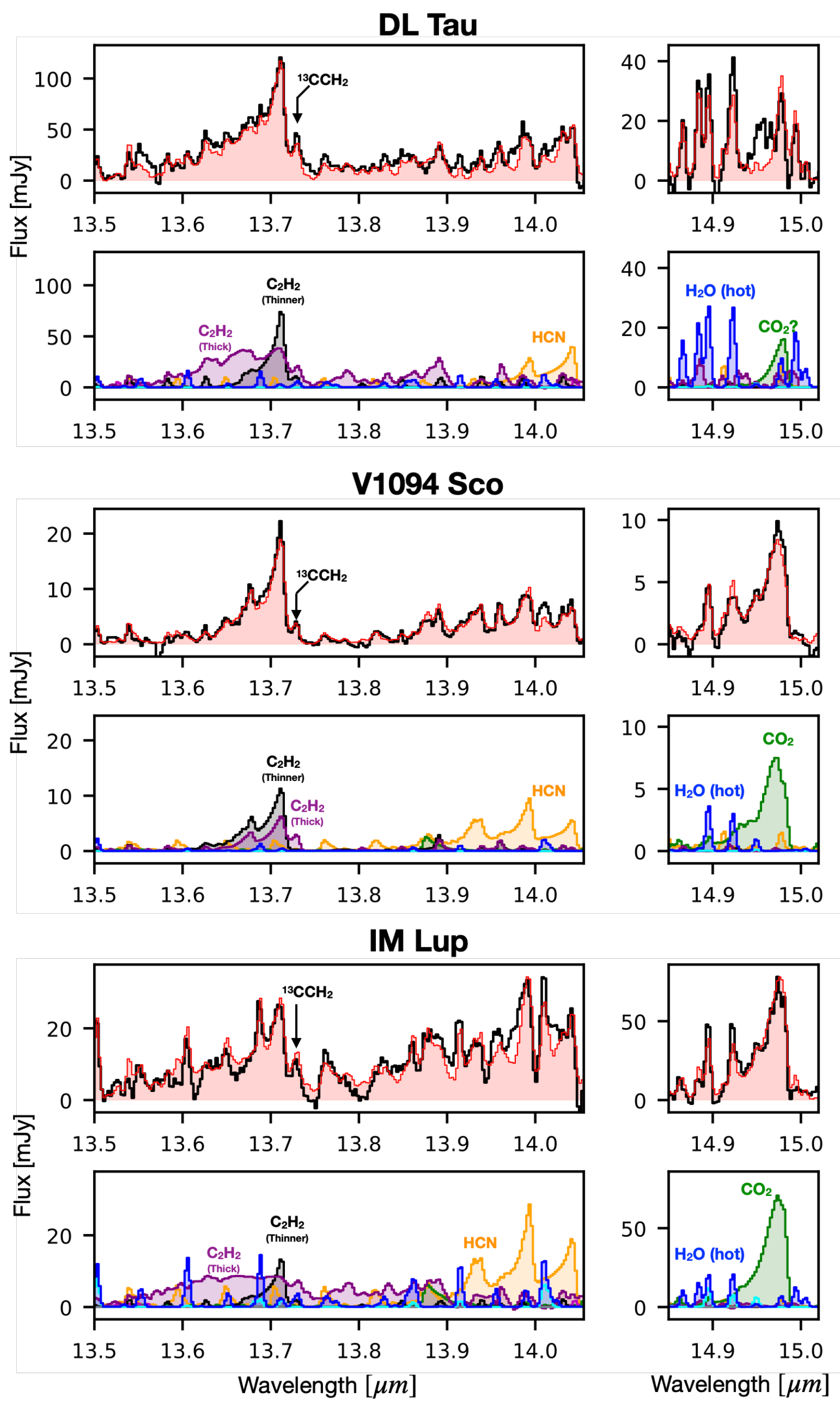}
\caption{Analysis of the 13.3-15~$\mu$m spectral range containing C$_2$H$_2$, HCN and C$_2$O features. Each panel shows the MIRI-MRS spectrum (black) and the total best fit model (red), and the contribution of each species. The best-fit slab model parameters are reported in Table \ref{table:best_fit_slab} and \ref{table:best_fit_slab2}. CO$_2$ is not securely detected in DL Tau and the slab model corresponds to the maximum emission of CO$_2$.}
\label{fig:C2H2_HCN_fit}
\end{figure}

\subsection{Acetylene (C$_2$H$_2$)}
The C$_2$H$_2$ feature does not vary only in strength but also in shape across the sample (Fig. \ref{fig:C2H2_HCN_fit}). In particular, \rev{the two sources with high C$_2$H$_2$/H$_2$O flux ratio} exhibit clear and prominent $Q$-branches \rev{whereas in IM Lup the Q-branch peaks at the same level as prominent nearby H$_2$O and HCN features in the spectral region shown in Fig. \ref{fig:C2H2_HCN_fit} (left panel).} 
Interestingly, the $^{13}$CCH$_2$ feature at 13.7 $\mu$m is detected, even in the oxygen-bright disk IM Lup. Overall, we find that the C$_2$H$_2$ feature cannot be adequately reproduced using a single slab model, in particular for DL Tau, \rev{as in VLMS where C$_2$H$_2$ is detected with a high SNR \citep{2023NatAs...7..805T,2024Sci...384.1086A}}. In particular, assuming a $^{13}$CCH$_2$:C$_2$H$_2 \simeq 35$, the  $^{13}$CCH$_2$ feature is systematically underestimated \rev{(see comparision in Fig. \ref{fig:compa_C2H2-one-vs-two})}. Adopting two slabs, we find very optically thick C$_2$H$_2$ in DL Tau and IM Lup with a small effective emitting area and column densities as high as $N($C$_2$H$_2) \simeq 4~\times 10^{19}$cm$^{-2}$ for IM Lup.  This value is at least two orders of magnitude higher than the values found from slab model fits of \textit{Spitzer}-IRS spectra, where C$_2$H$_2$ is found to be optically thin \citep{2011ApJ...731..130S}. \rev{We however stress that in IM Lup, the significant contribution of H$_2$O and HCN to the C$_2$H$_2$ along with the low signal-to-noise ratio and residual fringes (Fig. \ref{fig:C2H2_HCN_fit}, bottom left) challenge the fit of the feature and the exact best fit parameter of the optically thick component depends of the determination of the continuum. In V1094 Sco, a second slab allows to have a better fit of the weak $^{13}$CCH$_2$ $Q$-branch (see Fig.\ref{fig:compa_C2H2-one-vs-two}) 
} \rev{In DL Tau and IM Lup, the optically thick component is hotter and more compact that the optically thin component, suggesting a radial profile of C$_2$H$_2$ declining with radius reminicent of the components traced by H$_2$O \citep[see above and][]{2023A&A...679A.117G,2024ApJ...975...78R}.} The typical excitation temperature of $\simeq 200-500~$K derived here is lower by a factor of two compared to that inferred by \citet{2011ApJ...731..130S} using a large sample of \textit{Spitzer}-IRS observations or in the recent analysis of the Cycle 1 JDISCS sample by \citet{2025arXiv250507562A}. \rev{This difference might be due to the use of two slabs in our analysis, since the fit of the DL Tau feature by one slab gives a much higher temeprature of $900~$K (see Fig. \ref{fig:compa_C2H2-one-vs-two}).}  In any case, this demonstrates the importance of high spectral resolution and sensitivity to robustly infer basic parameters of the molecular features and the need for including multiple slabs when analysing the C$_2$H$_2$ feature. 
\rev{A consistent analysis of a large sample with fits of the C$_2$H$_2$ feature with more than one slab and proper accounting of H$_2$O and HCN contributions is required to confirm whether C$_2$H$_2$ is truly cooler in the three disks studied here than in other sources.}

\subsection{Water}

\begin{figure}
\centering
\includegraphics[width=0.5\textwidth]{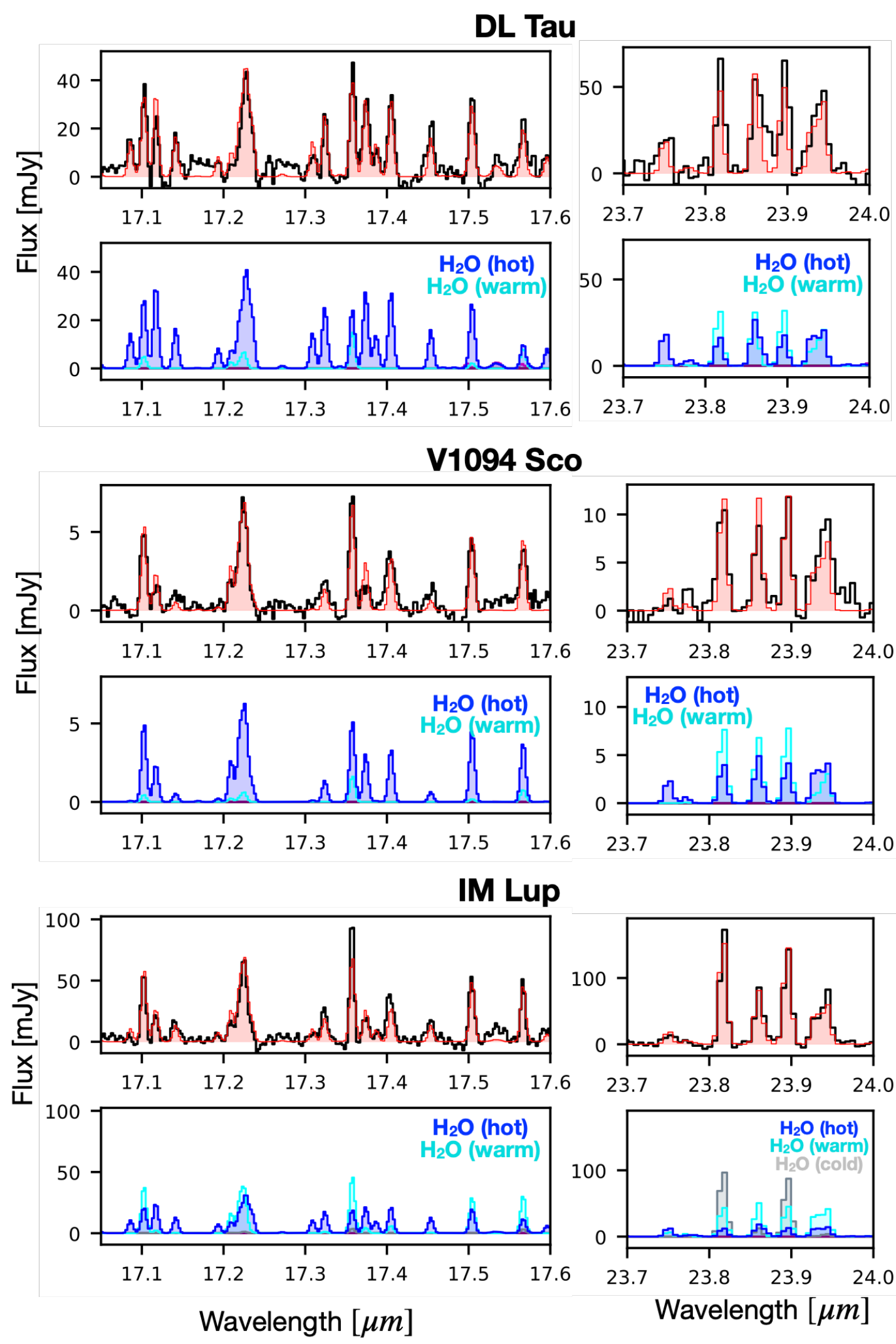}
\caption{Analysis of water emission in our sample. For each source, the observed JWST spectrum (black) is compared to the sum of all the best fit slab models (red). The total emission model is further split in the different slab components. The best-fit slab parameters are reported in Table \ref{table:best_fit_slab}. The water slab models are in blue, cyan, and gray in decreasing excitation temperatures. The water emission in DL Tau and V1094 Sco is well fitted by two slabs with a dominance of hot water. The IM Lup water spectrum is matched by three slabs.}
\label{fig:water}
\end{figure}

The pure rotational water lines observed by MIRI-MRS have been shown to trace a range of temperatures, with short-wavelength lines typically tracing hotter and long-wavelength lines tracing cooler water \citep{2023A&A...679A.117G,2023ApJ...957L..22B, 2024A&A...689A.330T,
2024ApJ...975...78R}. The water lines are detected in all three disks, but with a relatively low signal-to-noise and low line-to-continuum ratio in DL Tau and V1094 Sco. \rev{A fit with two slabs reproduces well the water emission of these two disks. For IM Lup, we find evidence for cold water} in the less excited lines \rev{(see Fig. \ref{fig:water}, bottom right panel)}, and fit the spectrum with three slabs. More sophisticated 1D models 
will be applied to our sample in a forthcoming publication (Temmink et al. in prep.). Figure \ref{fig:water}, \ref{fig:appendix-water}, and  \ref{fig:appendix-water2}  show that the water lines are well matched by our slab models. We recover that the overall spectrum traces primarily hot water around 700-900~K and, to a lesser extent, warm water around 300-400~K (see Table \ref{table:best_fit_slab}). In IM Lup, we further unveil a cold component around 200~K which is  visible in the water line complex around 23.8~$\mu$m (Fig. \ref{fig:water}, right panels, grey spectrum).

\subsection{Carbon dioxide} Carbon dioxide is detected in IM Lup and V1094 Sco (Fig. \ref{fig:C2H2_HCN_fit}) and the $Q$-banch is consistent with a relatively low temperature of 300-500~K, in line with other T Tauri disks \citep{2023A&A...679A.117G,2025arXiv250507562A} and somewhat warmer than in protostellar envelopes \citep{2024A&A...692A.197V}. The main $Q$-branch and the $P$- and $R$-branches are well reproduced by a single slab of gas. This suggests that CO$_2$ is emissive only under specific physical conditions, likely related to the competition between formation of H$_2$O versus CO$_2$ via hydrogen abstraction of OH (H$_2$+OH$\rightarrow$ H$_2$O+H versus CO+OH$\rightarrow$ CO$_2$+H), the latter being predominantly formed at low temperature \citep[$T\lesssim 300~$K,][]{2018A&A...611A..80B}. \rev{Indication of $^{13}$CO$_2$ is also seen in the residual spectrum of IM Lup confirming the detectability of the isotopologue for CO$_2$ dominated MIRI-MRS spectra \citep{2023ApJ...947L...6G,2025AJ....169..184S,2025A&A...693A.278V}.} In V1094 Sco, the column density is particularly high, as testified by the detection of the hot band \rev{around $16.2~\mu$m} (see Fig. \ref{fig:appendix-water}). In DL Tau, CO$_2$ is not clearly detected, and we provide in Table \ref{table:best_fit_slab2} an upper limit on its column density assuming the same emitting area and temperature as inferred for the warm water component.

\subsection{Other species}

\paragraph{C$_4$H$_2$} Diacetylene is detected in the C$_2$H$_2$-bright disk of V1094 Sco around 16~$\mu$m (see Fig. \ref{fig:appendix-water}), but remains undetected in DL Tau and IM Lup. This is the second detection of diacetylene in a disk orbiting a T Tauri star after DoAr 33 \citep{2024ApJ...977..173C}. The fit of the feature points toward a surprisingly low temperature of 150~K, just like in DoAr 33, with a rather small column density of $N=10^{17}~$cm$^{-2}$. Therefore, the observed C$_4$H$_2$ emission would originate from a different reservoir than C$_2$H$_2$, likely more extended as suggested by the best fit emitting radius of $R=2~$au which is a lower limit since the fit points toward optically thin emission. To obtain a constraint on the C$_4$H$_2$/C$_2$H$_2$ abundance ratio in the C$_2$H$_2$ emitting zone, we ran slab models assuming that C$_4$H$_2$ has the same temperature and emitting area as the two slabs of C$_2$H$_2$ and adjusted by eye the C$_4$H$_2$ column density to match the observed flux. This gives an upper limit of C$_4$H$_2$/C$_2$H$_2$<0.01 for V1094 Sco and an upper limit of C$_4$H$_2$/C$_2$H$_2$ $<0.005$ for DL Tau. Interestingly, these upper limits are well in line with the C$_4$H$_2$/C$_2$H$_2$ abundance ratio measured in the young protostellar system L1448-mm \citep{2024A&A...692A.197V}. However, this is much lower than the C$_4$H$_2$/C$_2$H$_2$ found in disks around very-low mass stars, where the ratios are 0.2-0.3 in J1605 and ISO-ChaI 147 \citep{2023NatAs...7..805T,2024Sci...384.1086A} when assuming that the optically thin C$_2$H$_2$ and C$_4$H$_2$ emission are collocated in these sources. 
\paragraph{HCN} Hydrogen cyanide is detected in all three sources. Although it is a carbon-bearing species, its emission is not significantly brighter in DL Tau and V1094 Sco than in IM Lup.
The results of slab fits point toward HCN residing in hot gas in V1094 Sco and IM Lup ($T \simeq 600-700~$K), in line with what is found in the majority of Class II disks \citep{2011ApJ...731..130S}. In contrast, the HCN feature in DL Tau traces cooler gas, around 400~K. This difference in excitation temperature is visible in the HCN spectra in Fig. \ref{fig:C2H2_HCN_fit} where the hot bands of HCN shortward of \rev{13.95}~$\mu$m are visible only in DL Tau and IM Lup.

\paragraph{CO} The tail of the CO rovibrational band is detected in all three sources, including the $v=1-0$ and $v=2-1$ transitions belonging to the $P$-branch in all three sources, except for V1094 Sco for which only the $v=1-0$ transitions are detected. We did not attempt to fit the CO emission with slab models because the lines covered by MIRI-MRS trace \rev{rotationally excited lines within $v=1$ vibrational state and more vibrationally excited levels which could trace a much higher temperature gas reservoir \citep[up to $\simeq 3,000~K$,][]{2024A&A...686A.117T} or are affected by optical depth \citep{2011A&A...533A.112H,2024A&A...683A.249F}.} 
A robust analysis of CO emission requires probing the full \rev{fundamental band}. We note that IM Lup was observed by CRIRES \citep{2013ApJ...770...94B}, but CO was not detected, highlighting the need for deep JWST/NIRSpec and future ELT-METIS observations.

\paragraph{OH} The rotational lines of OH are detected in all three disks with a similar line-to-continuum ratio. The high excitation temperature of 1,200-2,000~K suggests non-thermal excitation in all three sources and is consistent with findings for the majority of T Tauri disks. Because the 9-12 $\mu$m lines are not detected and based on the relative ratio within the quadruplets \citep{2024A&A...691A..11T}, the detected OH lines are unlikely to probe H$_2$O photodissociation but could be powered by chemical pumping: O + H$_2 \rightarrow$ OH + H \citep{2024NatAs...8..577Z}.

\paragraph{H$_2$} The rotational lines of H$_2$ are detected in all three sources from S(1) up to S(7). When compared to the continuum and the other molecular features, the H$_2$ lines do not appear to be dimmer in V1094 Sco, in contrast to other molecular features. This suggests that the H$_2$ emission does not originate from the inner disk but likely from the more extended disk surface and an outflow. This is consistent with the fact that the critical densities of H$_2$ lines are orders of magnitude lower than the other molecular emission detected by MIRI-MRS and can therefore trace more dilute and extended gas as found for a few other sources \citep[][]{2024ApJ...965L..13A,2025ApJ...980..148S}. In Appendix \ref{sec:appendix-H2-maps}, we show that extended emission of H$_2$ is detected in all three sources. The emission is systematically offset along the disk minor axis, supporting that H$_2$ lines trace outflows at least in DL Tau and V1094 Sco. Since no envelope is present, these are promising candidates for MHD disk winds where gas heating would be dominated by ambipolar diffusion \citep{2012A&A...538A...2P}. 

\subsection{Interpretation of the slab model parameters}

\label{subsec:over-sub}

\begin{figure}
\centering
\includegraphics[width=0.46\textwidth]{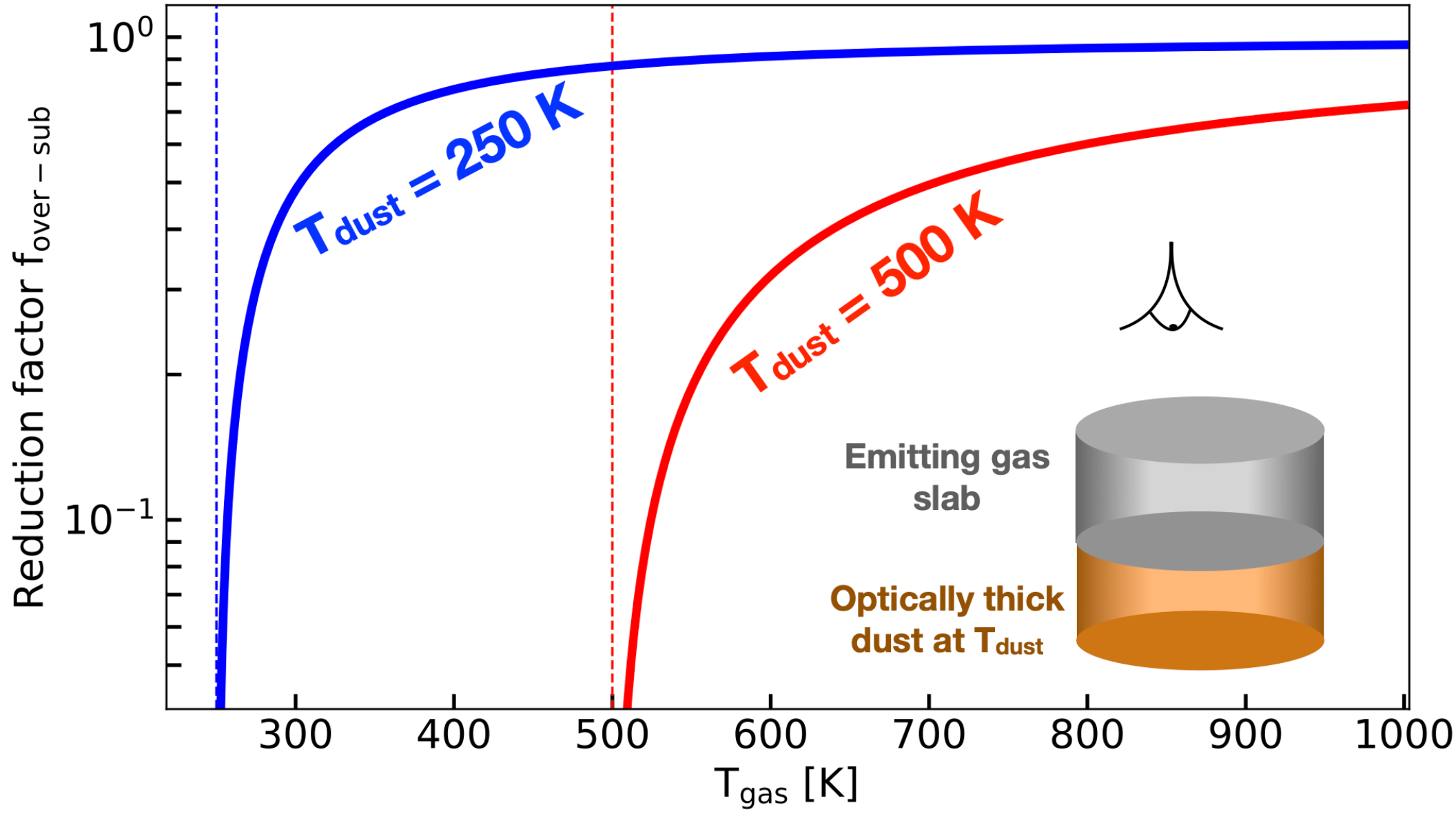}
\caption{Example of the reduction factor of a molecular feature seen in emission against optically thick dust as a function of the dust and gas temperature. Typical slab models used to analyze the molecular features neglect this effect by implicitly assuming a reduction factor of unity. In this example, we assume that the feature is at 15~$\mu$m. When the gas temperature is equal to the dust temperature, the feature vanishes. For gas temperature below the dust temperature, the molecular feature is seen in absorption.}
\label{fig:foversub}
\end{figure}

In this work, we analyze the molecular features using a gas slab model, as is commonly done in studies based on \textit{Spitzer}-IRS and MIRI-MRS data. However, the interpretation of the retrieved parameters remains difficult. A first effect, not taken into account in these models, and often confused with dust attenuation, is \textit{continuum over-subtraction} \citep{2021MNRAS.501.3427R,2022A&A...663A..58N}. When a molecular emission is seen on top of an optically thick dust continuum (see schematics in Fig.~\ref{fig:foversub}), the gas absorbs part of the continuum emission, reducing the amplitude of the gas features (see Appendix \ref{appendix:RT} for further details). As shown in Fig.~\ref{fig:foversub}, the reduction factor depends on the temperature difference between the dust photosphere and the excitation temperature of the gas.
Because the abolute flux in the gas slab models is proprotional to the effective emitting area, $\pi R_{\text{disk}}^{2}$, this quantity is overestimated by a factor $1/f_{\text{over-sub}}$. \rev{However, the spectral shape of the gas features is not affected (at least in narrow spectral regions). Therefore, the derived temperature remains unchanged, and the column density is likewise unaffected when the gas emission is optically thick.}

Which molecular features are expected to be the most affected by \textit{continuum over-subtraction}? If the emitting height is close to the mid-plane, the emitting region is more likely to be at a temperature close to that of the dust photosphere and therefore efficiently canceled by continuum over-subtraction. As a rule of thumb, a retrieved low temperature with a high column density is likely to emit deep down, where the gas temperature is close to the temperature of the dust photosphere. H$_2$O lines are likely not affected by continuum over-subtraction as they systematically trace warm and hot gas with relatively modest column densities. In contrast, the highly optically thick C$_2$H$_2$ features could trace regions more extended than the retrieved emitting area.

Another difficulty in slab model retrieval is to constrain the molecular abundance ratios. Depending on the energy involved in the observed transitions and the structure of the features themselves, features of two separate species can trace very different regions in the disk.
In the three disks studied here, C$_2$H$_2$, H$_2$O, and HCN have different emitting areas and temperatures (see Table \ref{table:best_fit_slab}), preventing us from deriving an estimate of abundance ratios. \rev{Ultimately, this is due to the complex 2D structure of protoplanetary disks. However, one can obtain upper limits of the C$_2$H$_2$/H$_2$O abundance ratio in each of the water reservoirs. To do so, we determine the maximum column density of C$_2$H$_2$ which remains consistent with the peak intensity of the observed C$_2$H$_2$ assuming the same emitting area and temperature as the considered water reservoir.
Since the features emit at similar wavelengths ($\simeq$ 13.5-18 $\mu$m), the impact of dust is similar, and the upper limit is robust and conservative. 
Following this method, we find that the C$_2$H$_2$/H$_2$O abundance ratio cannot be larger than $5 \times 10^{-2}$, $3 \times 10^{-2}$, and $5 \times 10^{-3}$ in the warm and hot water components in DL Tau, V1094 Sco, and IM Lup, respectively. The same methodology gives an upper limit on CO$_2$/H$_2$O of $5  \times 10^{-3}$, $1.5  \times 10^{-2}$, and $3  \times 10^{-2}$ in DL Tau, V1094 Sco, and IM Lup, respectively.}

\section{Discussion}
\label{sec:discussion}

Our analysis of the mid-IR molecular emission from three T Tauri disks with substantial reservoirs of cold pebbles reveals enhanced C$_2$H$_2$/H$_2$O flux ratios in DL Tau and V1094 Sco compared to IM Lup, and, more broadly, relative to \rev{the other full T Tauri disks of the MINDS sample (see Fig. \ref{fig:context})}. Yet, DL Tau and V1094 Sco are not deficient in O-bearing species with clear H$_2$O and OH emission and with even a prominent CO$_2$ feature in the latter. High C$_2$H$_2$/H$_2$O flux ratios \rev{in DL Tau and V1094 Sco} are also associated with a flat dust SED with the signature of silica grains (SiO$_2$).

In this section, we first discuss what local disk properties can drive the strength of the molecular emission, focusing on two possibilities: elemental abundances in the gas and dust optical depth effect. Assuming that the molecular emission is primarily indicative of the elemental abundance in the gas, we then explore two scenarios to account for the properties of our sample: inward radial transport of material from the cold outer disk and destruction of refractory carbon. We finally propose that silica grains could be related to the reformation of silicate in elevated C/O gas thereby providing a new diagnostic of C/O.

\subsection{What inner disk properties shape the molecular features?}

\subsubsection{Elemental abundances in the gas}

\label{subsec:elemental_abundance}
The determination of the elemental composition of the gas in relation to the interpretation of the atmospheric composition of gaseous planets, and in particular the C/O ratio, constitutes one of the driving motivations of JWST observations of disks. 
The most straightforward explanation for the high C$_2$H$_2$/H$_2$O flux ratio is a high gas-phase C/O in the inner disk. For C/O>1, oxygen is expected to be locked into CO, leaving a fraction of C in the gas-phase to form hydrocarbons. In fact, chemical models show that under irradiated and warm conditions, free carbon tends to be locked primarily into C$_2$H$_2$ instead of the thermodynamically more stable form: CH$_4$ \citep[][]{2011ApJ...743..147N,2022ApJ...934L..25D,2024A&A...681A..22K}. 

But gas-phase C/O is likely below unity even in DL Tau and V1094 Sco. The upper limits on C$_2$H$_2$/H$_2$O abundance in the water emitting region demonstrate that C$_2$H$_2$ is indeed underabundant compared to H$_2$O by at least of factor 20, in contrast with VLMS disks like J160532 \citep{2023NatAs...7..805T}. In addition, the OH emission that traces the top layers is not particularly dimmer in V1094 Sco and DL Tau when normalized to the continuum emission. In fact, it is well documented that a gas with a C/O ratio below unity produces C$_2$H$_2$ if it is irradiated by X-rays \citep{2013A&A...551A.118B,2015A&A...582A..88W,2026ApJ..1000..217D}. Below C/O<1, most of the carbon is indeed locked in CO but X-ray dissociate CO and produce free carbon atoms that can be converted in hydrocarbons. H$_2$O FUV shielding, effective for the water column densities inferred in all the disks studied here ($N($H$_2$O$)\gtrsim 10^{18}~$cm$^{-2}$), enhances these hydrocarbons chemistry thanks to a deficit in free atomic oxygen and OH in the deep layers \citep{2022ApJ...934L..25D,2022ApJ...934L..14C}. The same process would naturally explain the presence of an optically thick C$_2$H$_2$ reservoir in IM Lup in a gas that likely has a lower C/O than in DL Tau or V1094 Sco. 

The scenario of elevated (but below unity) C/O ratio is particularly promising in DL Tau, for which only weak H$_2$O is detected along with no clear sign of CO$_2$. Therefore, using C$_2$H$_2$/CO$_2$ flux ratio as a proxy for the C/O, as done by \citet{2025ApJ...978L..30L}, would lead to the same conclusions. In V1094 Sco, the $Q$-branch of CO$_2$ is relatively strong compared with H$_2$O. This is at odds with the thermochemical models of \citet{2018A&A...618A..57W} who find that CO$_2$ is severely quenched with elevated C/O ratio. 
Another solution to account for the presence of both CO$_2$ and C$_2$H$_2$ is a radial gradient in the C/O across the disk, as proposed by \citet{2023NatAs...7..805T} for J160532 \citep[see also][]{2026A&A...705A.222K}, \citet{2025A&A...694A.147G} for Sz 98, and \citet{2024ApJ...977..173C} for DoAr 33, or a small cavity \citep[$\lesssim 1~$au,][]{2024A&A...682A..91V}.
Yet, the C$_2$H$_2$/CO$_2$ flux ratio remains above the median value for the \citet{2025A&A...694A.147G} sample pointing toward elevated C/O. The detection of C$_4$H$_2$ in V1094 Sco, the second in a T Tauri disk, suggest active carbon chemistry likely to occur only for elevated C/O ratios. 
We also note that varying both C/H and O/H, \rev{Arabhavi et al. (2026) and Estève et al. submitted recently find that C$_2$H$_2$/H$_2$O flux is also enhanced for low O/H ratio. To match the flux ratio of DL Tau and V1094 Sco they increase the C/O ratio close to but below 1.}

\rev{To summarize, DL Tau and V1094 Sco have likely an elevated gas-phase C/O ratio, perhaps associated with low O/H compared with the other T Tauri disks of the MINDS sample but a C/O>1 is unlikely.} A robust estimate of the C/O ratio, which is beyond the scope of the present paper, needs to consider multiple tracers. 


\subsubsection{Dust opacity effect}

\rev{Gas, as traced by the C$_2$H$_2$/H$_2$O flux ratio, and dust, as traced by the silicate emission feature but also by the IR spectral index appear to be connected \citep[see also][]{2025arXiv250804692G}. For our selected sample of three disks}, DL~Tau and V1094~Sco show unusually shallow silicate features \rev{\citep[the weakest among the MINDS T Tauri disks,][]{2025arXiv250804692G} and exhibit the highest C$_2$H$_2$/H$_2$O flux ratios within the MINDS sample of full T Tauri disks. Within the MINDS sample,} this relationship is also recovered in CY~Tau \citep[see SED in][]{2025arXiv250515237T}. The strength of the silicate feature in these sources ranks among the weakest 5\% in typical T~Tauri disk populations \citep[][]{2009ApJS..180...84W}. 
The same correlation is also found in disks around very low-mass stars \citep{2025ApJ...984L..62A}.

A weak silicate feature indicates a disk with large silicate grains. This is further supported by the correlation \rev{found based on \textit{Spitzer} observations }between the strength of the silicate feature and its shape with weak silicate features corresponding to flatter features \citep{2009A&A...507..327O}. In addition, DL Tau, V1094 Sco, and CY Tau have a low mid-IR spectral index indicative of a well-settled and flat disk, possibly self-shadowed by their inner rim \citep{2004A&A...417..159D,2009A&A...501..383W}. \rev{This result is reminicent of the correlation between C$_2$H$_2$ luminosity and mid-IR spectral index found by previous \textit{Spitzer} and JWST studies based on a larger sample combining transition and full disks} \citep{2020ApJ...903..124B,2025arXiv250507562A}.

This connection between mid-IR molecular emission and mid-IR dust emission suggests that the elemental abundances in the gas are likely not the only parameter setting the observed molecular emission. In addition to variations in the elemental abundance effect, variations in the C$_2$H$_2$ flux could also originate from changes in the dust properties of the inner disk, which allow us to probe gas closer to the mid-plane. This is supported by the slab models \rev{showing high column densities of C$_2$H$_2$ which can be interpreted as infrared emission tracing deeper layers toward the mid-plane as proposed by \citet{2023NatAs...7..805T, 2025A&A...699A.194A}}. A reservoir of deep, highly optically thick C$_2$H$_2$ is supported by thermochemical models \citep{2018A&A...618A..57W}, in particular when the effect of UV shielding of H$_2$O is included \citep{2022ApJ...934L..25D}. The growth of dust grains would then lead to a decrease in the opacity of the grains, revealing the reservoirs of C$_2$H$_2$ normally hidden in other disks due to large amounts of small grains. This would explain why high column densities of both C$_2$H$_2$ and CO$_2$ are inferred in V1094 Sco. The water mid-IR emission would be less affected by a decrease in the opacity of the dust because water spectral features are made of isolated lines and blends that easily saturate at high column densities \rev{\citep[see Fig. 7 in][for an illustration]{2025ApJ...984L..62A}}. This contrasts with C$_2$H$_2$, which has a myriad of very weak lines which start to emit at high column densities. The same holds to a lesser extent for CO$_2$ and HCN.

The key parameter in this context is the location at which dust emission becomes optically thick around 12-16~$\mu$m where the major O- and C-carriers are emissive. For a given vertical dust column density and assuming for the sake of simplicity a mono-disperse population, increasing the dust grain size from 0.1 to 5 $\mu$m typically reduces the opacity at 13~$\mu$m by a factor of 5 to 10, thereby exposing larger column densities of C$_2$H$_2$. The vertically integrated dust-to-gas ratio could also be significantly smaller than the ISM value due to e.g., pebble traps outside of the inner disk. However, this reasoning assumes that the molecular layer remains unchanged as the dust evolves. This is a significant caveat, since grain growth or depletion reduces the H$_2$ formation rate and increases UV penetration, thereby shifting the molecular reservoir to lower altitudes. Consequently, chemical feedback is expected to limit the effect of reduced mid-infrared opacity on the C$_2$H$_2$ or CO$_2$ emission. Also, the brightness of molecular features is not merely set by dust opacity but also by the temperature gradient between the emitting gas layer and the dust photosphere as discussed in Sec. \ref{subsec:over-sub}.

Several thermochemical models have included grain growth and settling. For example, \citet{2019A&A...626A...6G} and \citet{2023A&A...672A..92A} explored how the mid-IR emission responds to the dust properties finding an overall increase in molecular emission for dust-depleted inner disks or for increased dust size. However, as demonstrated by \citet{2022ApJ...934L..25D}, FUV shileding of the gas plays a central role in the chemistry of inner disks, notably for C$_2$H$_2$ and HCN, and to a lesser extent CO$_2$. In fact \citet{2023A&A...672A..92A} neglect FUV shielding by H$_2$O and do not predict detectable amounts of C$_2$H$_2$. Therefore, further thermochemical models that explore the combined effects of dust properties and FUV gas shielding would be valuable. Moreover, our results highlight the usefulness of jointly analyzing the continuum emission and molecular features, and point to the potential of specific molecular tracers that probe regions closer to the disk mid-plane.

In summary, the strength of the molecular features is likely the result of a combination of dust optical depth effect and elemental abundances. Only detailed thermochemical models coupled to radiative transfer can uniquely distinguish these effects. In the following of the discussion we assume that the molecular features are primarily a tracer of the elemental abundance and explore two scenarios accounting for the diversity of the mid-IR emission: radial transport of elements and destruction of refractory carbon.

\begin{figure}
\centering
\includegraphics[width=0.45\textwidth]{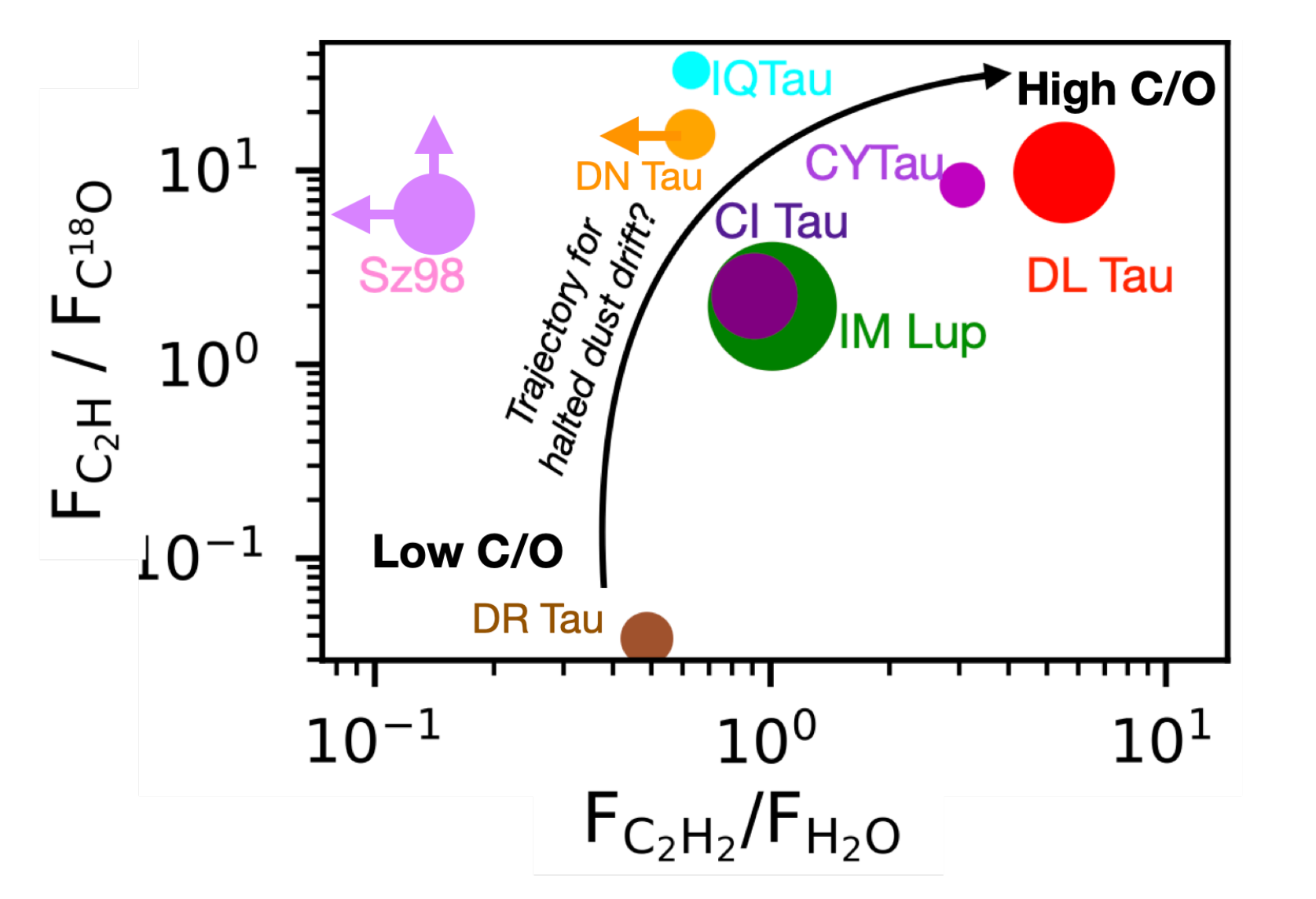}
\caption{Comparison between the composition of the inner versus outer disks focusing on disks with a significant amount of pebbles in their outer disk ($M_{\text{Pebble}}\gtrsim 25~M_{\oplus}$). The size of the dots represents the size of the disk as measured in $^{12}$CO(2-1) emission. \rev{The millimeter line fluxes are described in \ref{table:appendix-C2H} and the mid-IR are from \citet{2025arXiv250804692G,2025A&A...694A.147G}.}}
\label{fig:inner-vs-outer}
\end{figure}

\begin{figure*}
\centering
\includegraphics[width=0.9\textwidth]{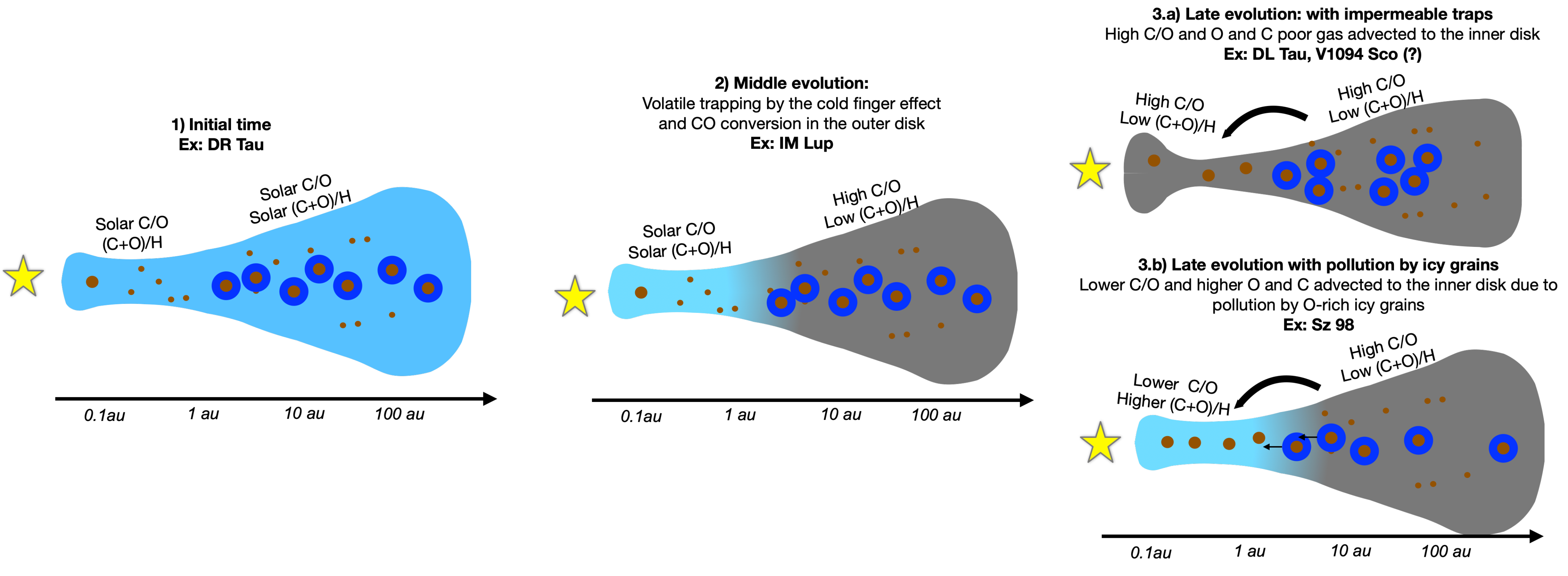}
\caption{Scenario to account for the chemical composition of inner disks under the paradigm of radial transport of volatiles. The annotated C/O refers to the gas. If icy pebbles are retained in the outer disks, the composition of the inner disk is regulated by the composition of the gas from the outer disk. Due to the time evolution of the outer disk and the advection of the gas toward the inner disk, a mismatch in C/O in the gas is expected. This mismatch is expected to be reduced with time. However, a small fraction of drifing icy grains can contaminate the flow and reduce the C/O ratio in the inner disk compared with the outer disk. This scenario is further quantified by a toy model (see Fig. \ref{fig:toy-model-outer-inner} and Appendix \ref{appendix:radial_transport}).}
\label{fig:scenario-schematics}
\end{figure*}

\subsection{First scenario: radial transport from the outer disk}

\label{subsec:inner-vs-outer-transport}

Over the past decade, the radial redistribution of the species from the cold outer disk to the warm inner disk has emerged as a key process setting the elemental abundances in disks \citep[e.g.,][]{2017MNRAS.469.3994B,
2020ApJ...899..134K}. This includes the possibility of fast radial drift of icy pebbles bringing a large amount of oxygen that sublimates into the gas-phase or their retention in pressure maxima outside of the snowlines of the main O- and C-bearing species with advection of gas from the outer disk. The giant disks studied here are unique sources to test the radial transport of elements because their substructures, clearly imaged by ALMA, are pebble traps \citep[][]{2023ASPC..534..423B}. 
Therefore, they are the best candidates for halted pebble drift with a large reservoir of icy pebbles retained in the outer disk and replenishment of the inner disk by gas with high C/O flowing from the outer regions.

\subsubsection{Link between the gas in the inner versus outer disk}

Can the radial transport of volatile species from the outer to the inner disk produce elevated C/O in the inner disk when pebble drift is quenched? To test this hypothesis, we first collected constraints on the composition of the gas in the outer disk using the sub-millimeter lines of C$_2$H (3-2) and C$^{18}$O (2-1) as a proxy for the gas-phase C/O ratio in the outer disk. To put our sample in context, we also include all the Class II disks with measured  C$_2$H (3-2) and C$^{18}$O (2-1) line fluxes by NOEMA, ALMA, and SMA, and published MIRI-MRS data. These sources are also among the 25\% brightest disks in pebble emission ($M_{\text{dust}}\gtrsim 25~M_{{\oplus}}$) and are therefore likely in the same regime of halted pebble drift as our sample of giant disks. Unfortunately, no published C$_2$H data are available for V1094 Sco. The millimeter data are complemented by unpublished data from the NOEMA-PRODIGE program (Semenov et al. in
prep). The references to the data are described in Appendix \ref{sec:appendix-C2H-C18O-flux-measurments}. 

Figure \ref{fig:inner-vs-outer} presents the C$_2$H/C$^{18}$O flux ratio from the outer disk versus the C$_2$H$_2$/H$_2$O flux ratio from the inner disks for eight T Tauri disks. Detailed chemical modelling is required to infer the exact value of the C/O ratio in the gas from the line ratio. In this discussion, we assume that the C$_2$H/C$^{18}$O and C$_2$H$_2$/H$_2$O  flux ratio are proxies for the C/O ratio in the gas \rev{(see Sec. \ref{subsec:elemental_abundance} for the latter)}. \citet{2019A&A...631A..69M} showed that C$_2$H becomes much brighter for C/O above 1 but also depends on disk mass. By rescaling the C$_2$H flux by C$^{18}$O, we mitigate dependency on the disk mass. In fact, the C/O ratio have been estimated by detailed modelling for DR Tau and IM Lup finding values of, respectively, $\simeq 0.5$ \citep{2024ApJ...973..135H} and $\simeq 0.8$ \citep{2018ApJ...865..155C}, and constrained by standard disk models for Sz 98 finding C/O>1 \citep{2019A&A...631A..69M}. This ordering in the inferred C/O is recovered in Fig. \ref{fig:inner-vs-outer}, supporting our qualitative approach. Regarding the C$_2$H$_2$/H$_2$O flux ratio, existing chemical models show that this line ratio correlates with the gas-phase C/O ratio \citep{2018A&A...618A..57W,2021ApJ...909...55A} even though the total (C+O)/H abundance is expected to contribute (Arabhavi et al. 2026, Estève et al. in prep.).

Figure \ref{fig:inner-vs-outer} shows that no clear correlation is seen between the inner versus outer disk molecular emission. Most of the disks have high C$_2$H/C$^{18}$O flux ratio, pointing toward an elevated C/O ratio in their outer regions as found in surveys \citep{2019A&A...631A..69M,2019ApJ...876...25B}. This contrasts with the C$_2$H$_2$/H$_2$O flux ratio, which is high only for DL Tau and, to a lesser extent, for CY Tau. Therefore, a mismatch between the outer carbon-rich disks and the inner oxygen-rich disks is found for the bulk sample. In this context, DL Tau appears to be an outlier in terms of a C-rich inner disk in addition to having a C-rich outer disk. IM Lup is in between, with a better match between the inner and outer disk compared with the rest of the sources. The other giant disk, Sz 98, is an extreme example of an inner disk very bright in O-bearing species and an outer disk very bright in C-bearing species \citep{2023A&A...679A.117G}. DR Tau is another interesting source with a relatively low C$_2$H$_2$/H$_2$O \citep[][]{2024A&A...686A.117T} and weak C$_2$H \citep{2024ApJ...973..135H}, making it the other source with a possible match between inner and outer composition.

\subsubsection{Toy model for halted pebble drift}

Dust and gas can transport material from the outer to the inner disks. This happens on very different time scales (drift vs. viscous or wind timescale). Gaps can change the flow structure both for the gas and the pebbles. The exact effect of the gaps depends on the mechanism which produced them \citep[e.g.,][]{2024A&A...691A..72L} and the time when the gap was created. \citet{2025A&A...694A.147G} discussed the scenario of partially halted icy-pebble drift assuming that the outer disk is carbon rich. The mismatch in terms of inner versus outer disk composition is then interpreted as the result of leaky dust traps supplying the inner disk with of O-rich ices. We update the scenario of \citet{2025A&A...694A.147G} in Figure \ref{fig:scenario-schematics} by stressing the importance of the chemical timescale required to convert CO into less volatile species (CO$_2$, H$_2$O, CH$_3$OH) and increase the C/O in the gas phase. According to this scenario, the overall disk would start oxygen-rich. DR Tau would be a representative source of this early stage, likely due to a streamer feeding its outer disk in pristine O-rich material \citep{2023ApJ...943..107H}. The outer disk would then convert CO into CH$_4$ and less volatile species like CO$_2$, CH$_3$OH, H$_2$O decreasing the (C+O)/H and increasing C/O in the gas over a timescale of ~1 Myr \citep[Sturm et al. 2022,][]{2020ApJ...891L..17Z}. This would result in a mismatch between the inner and outer disk C/O ratio, as seen in the disk of IM Lup. It is then only at a later stage that the gas reaches the inner disk, leading to an overall carbon-rich disk. DL Tau would be the best example of this evolved stage. However, in the presence of a small leakage of icy-grains, the increase of C/O in the inner disk is severely limited because the H$_2$O and CO$_2$ rich, low C/O, icy grains are the main carrier of oxygen and carbon. This could explain the mismatch between inner and outer disk seen in Sz 98 by \citet{2023A&A...679A.117G} and to lesser extent DN Tau and IQ Tau (Fig. \ref{fig:inner-vs-outer}).

\begin{figure}
\centering
\includegraphics[width=0.45\textwidth]{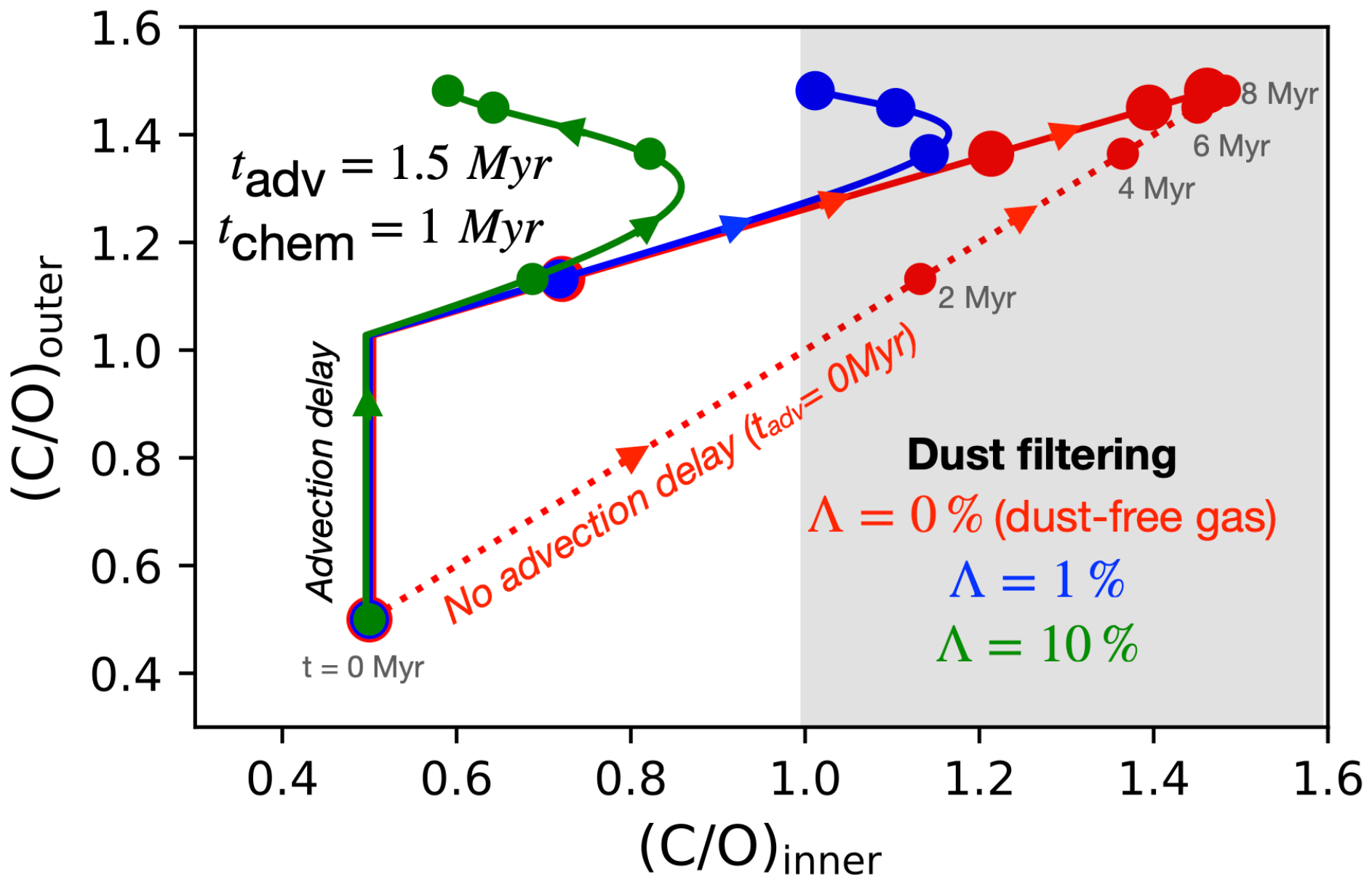}
\caption{Composition of the outer versus inner disk predicted by our toy model of halted pebble drift. The C/O ratios in the gas of the outer versus inner disk are indicated every 2~Myr starting from solar values. Increased C/O ratio in the inner disk is achieved following the increase in C/O ratio in the outer disk and the subsequent advection of the gas. This gas advection introduces a lag between the outer disk versus the inner disk composition. The effect of pollution by a small fraction of ice-coated grains is also shown. Because for $t\gtrsim t_{\text{chem}}$ ice carries most of the oxygen and carbon, they have a pronounced effect on the inner disk, limiting the increase in C/O by bringing low C/O material.}
\label{fig:toy-model-outer-inner}
\end{figure}

To better quantify the scenario of radial transport of volatiles, we build a toy model aimed at capturing the essential aspects of our scenario (see Appendix \ref{appendix:radial_transport} for more details). \rev{This model applies to disks that retain a significant amount of pebbles.} We model the disk as two reservoirs: a cold outer reservoir containing gas and icy grains, and a warm inner reservoir fed by the advection of gas and ice from the outer disk. The evolution of the C/O and O/H of the gas in the outer disk is prescribed using observational constraints gathered by \textit{Herschel} and ground-based millimeter observations. Specifically, the O/H decreases down to O/H $= 3 \times 10^{-6}$ and the C/O increases up to C/O=$1.5$ over a chemical timescale $t_{\text{chem}} = 1~$Myr (see left panels Fig. \ref{fig:appendix-toy-model-inner-vs-outer-O-H}). The composition of the advected material is calculated by considering pollution by oxygen-rich ice parameterized by $\Lambda$, the gas-to-dust mass ratio of the advected material with respect to the ISM value. Furthermore, the time required for the material in the outer disk to advect to the inner disk is modelled by a delay of $t_{\text{adv}} = 1.5~$Myr. \rev{A similar approach is also followed by \citet{2025ApJ...990L..72K} though their delay corresponds to the travel time of large pebbles.} This value is justified by the fact that the advection timescale is about the timescale associated with the transport of angular momentum, which is found to be typically of about 0.5-2~Myr \citep{2017MNRAS.472.4700L,2022MNRAS.512L..74T,2025ApJ...989....7T}. Our toy model is reminiscent of the more sophisticated model of \citet{2025arXiv250711631S}, which also takes into account the chemical conversion of CO \rev{(also referred to as CO missing problem or carbon depletion)}, a key process that remains disregarded in other radial transport models.

We show in Fig. \ref{fig:toy-model-outer-inner}~the predicted evolution of the outer versus inner disk composition. The case of $\Lambda=0$ (red solid line) corresponds to a complete blocking of icy-grains (either pebbles or micron-size grains). The C/O in the outer disk first increases as the outer disk chemistry proceeds. After an advection timescale $t_{\text{adv}}$, the increased C/O propagates to the inner disk, forming a turning point with a straight trajectory until both the outer and inner disk reach the final C/O of 1.5. We find that the chemical and advection timescales need to be typically shorter than few times the age of the system to reach steady state. The fully halted pebble drift can qualitatively account for the overall pattern in  distribution of the sources in the C$_2$H$_2$/H$_2$O versus C$_2$H/C$^{18}$O plane shown in Fig. \ref{fig:inner-vs-outer}.

However, it is unlikely that the icy grains are fully blocked in pressure traps. \rev{The efficiency of pebble traps depends on their depth but also distance from the central object as suggested from the correlation between the location of the inner substructure and the presence of cold water excess \citep{2025ApJ...990L..72K}.}  Micron-size grains resulting from the collisional cascade of pebbles are well coupled to the gas and provide a threshold value for the leakage of the traps \citep{2018ApJ...854..153W}. These small grains bearing ices are evidenced by JWST IFU observations of edge-on disks \citep[e.g.,][]{2023A&A...679A.138S,2025A&A...698A...8D,2026arXiv260318163B} but their absolute abundance remains to be quantified to evaluate the mass of solids flowing across pebble traps. For $\Lambda = 0.1$, corresponding to retention of $90\%$ of the dust (compared to ISM), Fig. \ref{fig:toy-model-outer-inner} shows that the advected gas remains well below the C/O of the outer disk. This significant effect of a tiny amount of icy-grains is due to the decline of O/H and C/H in the gas of the outer disk down to the values estimated from \textit{Herschel} and ALMA observations of Class II disks \citep[$10^{-2}-0.1$,][]{2011Sci...334..338H,2017ApJ...842...98D,2016A&A...592A..83K,2017A&A...599A.113M,2022A&A...660A.126S}. In fact, as time proceedes, O and C get locked into icy grains, enhancing their impact on the inner disk. Quantitatively, we find that a blocking efficiency of about 95\% ($\Lambda = 5 \times 10^{-2}$) is required to get C/O about unity for few Myr. 

The requirement of very efficient dust retention is likely the reason why disks that have retained a large reservoir of grains and have reached a high C/O ratio in their outer disk do not systematically have high C$_2$H$_2$/H$_2$O flux ratios. DL Tau would then be a special case of efficient retention of icy dust and well evolved outer disk. In fact, no excess of cold water is seen, suggesting the absence of substantial sublimation of drifted water-rich pebbles in DL Tau. In addition, detailed analysis of its outer disk confirms an extreme C and O depletion in the gas-phase \citep{2022A&A...660A.126S}. However, full blocking is unlikely since our analysis of DL Tau shows that C/O in the inner disk is likely elevated but below 1. For the other extreme, Sz98 would be a disk dominated by the pollution of O-rich ice as proposed by \citet{2023A&A...679A.117G}. For IM Lup, the inner disk could either be polluted by drifting pebbles, as suggested by the excess of cold water and by the prominent CO$_2$ feature \citep[see discussion in][for CX Tau]{2025A&A...693A.278V}, or be O-rich due to the early evolutionary stage of its outer disk with a rather small but super-solar C/O ratio \citep{2018ApJ...865..155C}. A filtering efficiency of about $95\%$ would enhance the total gas-to-dust ratio in the inner disk by a factor 20, \rev{which could} reduce the optical depth of the disk and produce a flatter SED as seen in the models \citep[e.g.,][]{2019A&A...626A...6G}.

We stress that our scenario should be contrasted with the scenario often proposed to account for the presence of C-rich gas around very low-mass stars (VLMS). In fact, \citet{2025ApJ...978L..30L} and \citet{2025A&A...699A.194A} find an anti-correlation between the dust mass and the column density of C$_2$H$_2$ (estimated from the C$_2$H$_2$/$^{13}$CCH$_2$ ratio) which contrasts with \rev{the high C$_2$H$_2$/H$_2$O flux ratio observed in two of the three most pebble-rich T Tauri disks  of the MINDS sample ($M_* > 0.4~M_{\odot}$)}. In disks around VLMS, it is proposed that dust drift is not halted but, on the contrary, accelerated, leaving behind an O-poor inner disks \rev{and pebble mass in their outer regions} after a brief avalanche of icy pebbles \citep{2023A&A...677L...7M,2025arXiv250711631S}. In summary, two pathways can lead to a C-rich and O-poor inner disk: halted pebble drift, more likely to occur in massive disks around T Tauris, and accelerated pebble drift, more likely to occur around VLMS.

\subsection{Other scenario: destruction of refractory carbon}
\label{subsec:C-destruction}
 
It was recently proposed that high C$_2$H$_2$/H$_2$O flux ratio traces the destruction of carbonaceous grains, which contain roughly half of the overall carbon budget \citep{2023NatAs...7..805T,2024ApJ...977..173C}, with important implications for the composition of exoplanets \citep{2023ApJ...949L..17B}. Starting from a solar C/O ratio of about 0.4, destruction of refractory carbon would increase the C/O by about a factor of 2, bringing it close to, but still below, unity, consistent with our abundance constraints in DL Tau and V1094 Sco. Thermal sublimation of refractory carbon at relatively low temperatures of 300-500~K would make this process effective in inner disks \citep{2021SciA....7.3632L} but requires a heavily processed form of carbon grains. Carbon grains could also be destroyed by photolysis, oxidation, or reaction with atomic hydrogen \citep{1995ApJ...447..848L,2014A&A...569A.119A,2017A&A...606A..16G,2025A&A...694A..89B}. Regardless of the exact destruction process, one of the main conundrums is why in some specific disks, carbon grain destruction would lead to observational fingerprints (high C$_2$H$_2$/H$_2$O flux ratio) whereas in the majority of T Tauri disks it would remain invisible.        

\citet{2024ApJ...977..173C} argue that elevated C / O in the gas triggered by sublimation around 300-500~K can be maintained if the accretion rate is low, allowing gaseous carbon to linger in the inner disks. Our results are in contrast with the lingering carbon scenario because DL Tau and V1094 Sco have accretion rates of few $10^{-8}~M_{\odot}$ yr$^{-1}$, slightly larger than the average accretion rate of 0.8~$M_{\odot}$-mass T Tauri stars \citep{2023ASPC..534..539M}, and more than two orders of magnitudes above DoAr 33 (Table \ref{table:source_properties}). The proposal by \citet{2024ApJ...977..173C} highlights the importance of the transport of the gas within the inner disk following the destruction of carbonaceous grains. In particular \citet{2025A&A...699A.227H} demonstrate that gas in the inner $\simeq 10$~au can be significantly enhanced in carbon following the destruction of grains in the innermost regions. In DL Tau and V1094 Sco, the outward radial transport of sublimated carbon grains, mediated by viscousity, could be enhanced due to for example low Schmidt numbers. This outward transport is particularly crucial here to account for the presence of C$_4$H$_2$ at a temperature much lower than the minimum sublimation temperature of refractory carbon (150~K versus 300~K). 
\rev{However, this scenario does not explain why in the MINDS sample, the two  T Tauri disks with the higest C$_2$H$_2$/H$_2$O flux ratios are also the largest and the most massive in pebbles.}
On the contrary, halted pebble drift would rather quench the avalanche of solid carbon that is required to sustain elevated C/O in the inner disk by carbon grain destruction \citep{2025A&A...699A.227H}. Therefore, destruction of refractory carbon, though likely occurring in inner disks, does not appear as a compelling scenario to account \rev{for the elevated C$_2$H$_2$/H$_2$O flux ratio} found in DL Tau and V1094 Sco.
 
\begin{figure*}
\centering
\includegraphics[width=0.9\textwidth]{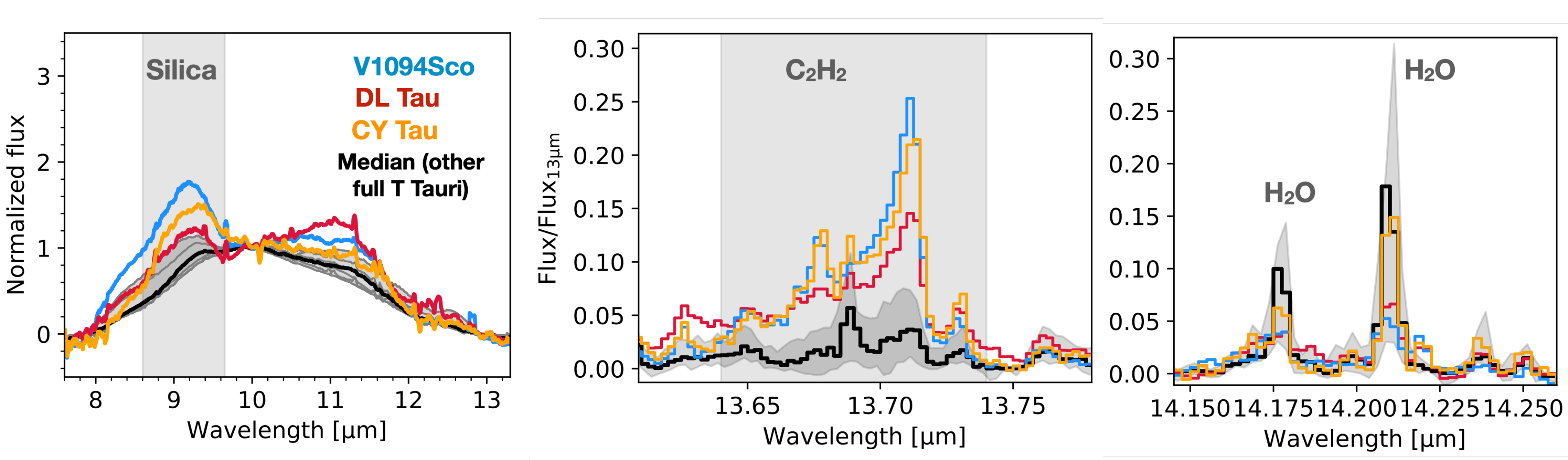}
\caption{Evidence for a link between dust stoichiometry and molecular emission. \textit{Left}: continuum subtracted spectra normalized to the continuum flux at 13~$\mu$m of the four sources showing the higest C$_2$H$_2$/H$_2$O flux ratios versus the full T Tauri disk of the MINDS sample (median spectrum in black and individual spectrum in grey). DL Tau, V1094 Sco, and CY Tau show a prominent silica feature at $9~\mu$m that is not present in IM Lup. \textit{Middle and right}: the former disks exhibit also a prominent C$_2$H$_2$ compared with H$_2$O.}
\label{fig:silica}
\end{figure*}


\subsection{First evidence for a connection between dust stoichiometry and gas emission}

The silicate emission feature also encodes information about the grain composition. In DL~Tau and V1094~Sco, a distinct feature at 9\,$\mu$m reveals the presence of silica, previously detected in approximately 10\% of protoplanetary disks by \textit{Spitzer}. This relationship extends to other high C$_2$H$_2$/H$_2$O disks: CY~Tau, a MINDS source (see Fig. \ref{fig:silica}) and DoAr~33 a JDISCS source, both showing a clear silica feature. To our knowledge, this is the first time a link between dust stoichiometry and molecular emission is identified, albeit in a small sample. \rev{In a submitted paper, we further quantify this relationship, and show that it is robust when using the retrieved mass fraction of silica from dust model fitting (Varga et al. submitted).} The origin of this correlation remains uncharted. Silicate grains in disks are known to undergo annealing and crystallization when exposed to high temperatures. The mid-IR signatures of reprocessed silicates imply an outward transport of grains from high temperature sites to the 0.5-10~au regions where they emit \citep{2009A&A...507..327O,2024A&A...687A.275J}. Since silica is not present in the interstellar medium \citep{2004ApJ...609..826K}, its presence must result from in-situ reprocessing. Silica can be produced via annealing of silicate dust \citep{2000A&A...364..282F} but there is no obvious reason why the annealing depends on the C/O ratio in the gas. 

Prominent silica features are observed in some debris disks and are interpreted as the recondensation of vaporized bodies following giant impacts \citep{2009ApJ...701.2019L}. In gas-rich disks, giant impacts are unlikely. However, silicates are also formed at the sublimation front from SiO vapor ($\simeq 0.05-0.1~$au for T Tauri disks), as evidenced by \citet{2025Natur.643..649M} where both crystalline silicate grains and perhaps silica are detected along with hot SiO gas in the inner region of a young protostellar disk. The silica features observed in high C$_2$H$_2$/H$_2$O flux ratio disks could then be the result of the recondensation of silicates in an oxygen-poor gas at the sublimation front of silicate followed by outward radial transport. This is well in line with out-of-equilibrium calculation of dust formation in AGB winds where silica features are predicted to become more prominent than olivine and forsterite as the C/O approaches unity \citep{2002A&A...382..256F}. This further supports our claims that in DL Tau and V1094 Sco, C/O ratio in the gas is close to, but not above 1, otherwise dust formation would lead to SiC grains rather than silica. In any case, the discovery of a correlation between gas and dust emission in T Tauri disks open new avenues to build a consistent picture of disk evolution coupling dust and gas composition.

\section{Conclusions}
\label{sec:conclusions}

In this study, we investigate the mid-infrared emission from the three largest and pebble-rich disks orbiting single T Tauri stars, selected from the MINDS program. These disks are potential progenitors of systems hosting giant planets on wide orbits. They are also expected to be the best example of halted pebble drift, where the composition of the inner disk is influenced by dust traps. Our analysis of these three giant disks is complemented by previously published JWST data on other disks, as well as by constraints on the gas composition in their outer regions. Our conclusions are:


\begin{itemize}
    \item The inner regions of the \rev{three giant disks of the MINDS survey} exhibit a great diversity in terms of molecular emission despite the very similar properties of their outer disk and of their central star\rev{, in line with other studies \citep[e.g.,][]{2025arXiv250507562A}}. Of the three sources studied here, DL Tau and V1094 Sco show the highest C$_2$H$_2$/H$_2$O flux ratios \rev{among the MINDS sample of full disks orbiting single T Tauri stars.}

     \item \rev{Slab model fits allow us to place conservative upper limits on the C$_2$H$_2$/H$_2$O and CO$_2$/H$_2$O abundance ratio in the H$_2$O emitting regions. Even in DL Tau and V1094 Sco, the C$_2$H$_2$/H$_2$O abundance ratio is below $5 \%$. This suggests that} in these two sources, the volatile C/O in the inner disk, though likely supersolar, is smaller than unity since C$_2$H$_2$ is predicted to be one of the main C carrier in high C/O gas. Therefore, C/O>1 might be very rare in the inner regions of T Tauri disks. \rev{This proposal is also supported by on-going modelling efforts which directly predict the molecular line fluxes (Arabhavi et al. 2026 and Estève et al. submitted).}
    \item We unveil a mismatch between the outer carbon-rich and inner oxygen-rich regions in a sample of eight T Tauri disks that host substantial amounts of pebbles. 
    Only DL Tau and another source of the MINDS program, CY Tau, are bright in hydrocarbons in both the outer and inner disks. We propose that the retention of oxygen-rich icy grains in the outer disk and the advection of gas with high C/O drives an increase in C/O and a decrease in the (C+O)/H ratio in the inner disk, resulting in prominent C$_2$H$_2$ emission. DL Tau and V1094 Sco would be the best instances of this category of disk. Sublimation of refractory carbon is a less compelling scenario to account for bright C$_2$H$_2$ because it requires lower accretion rates than observed and efficient dust radial drift.
    \item With a toy model, we demonstrate that a match between a C-rich gas composition of the inner versus a C-rich outer disk requires: (1) Efficient blocking of icy grains of about $\simeq 95 \%$ to obtain C/O $\simeq 1$, and (2) an age that is long enough to chemically process the outer disk and advect the high C/O gas from the outer to the inner disk. 
    \item Therefore, inner disks are unlikely to reach C/O above unity, even when the majority of the solids is blocked in the outer disk. This is because as the outer disk C/O increases with time, the oxygen-rich icy grains lock most of the carbon and oxygen. A small amount of icy grains brought, for example, by small grains limits the incease in C/O. IM Lup and Sz 98 would be instances of polluted inner disks as hinted by their excess of cold water, \rev{though the outer disk of IM Lup might also be not evolved enough to have reached high C/O in the gas}. This highlights the importance of considering the CO chemical conversion in models of radial transport of volatiles.
    \item Silicate grains are found, for the first time, to exhibit a prominent silica (SiO$_2$) component in sources with a high C$_2$H$_2$/H$_2$O flux ratio. \rev{This is recovered in the three T Tauri disk of MINDS with the highest C$_2$H$_2$/H$_2$O flux ratios: DL Tau, V1094 Sco, CY Tau. We also note that DoAr 33, observed by the JDISCS collbaoration has also this coincidence between silicate feature and molecular emission.} We propose that silica is formed at the sublimation front of the silicates in a gas with super-solar but below unity C/O ratio. This consititutes a new promissing tracer of the C/O in planet-forming disks. 
\end{itemize}


MIRI-MRS provides unique data to understand how the diversity of exoplanets emerges. In T Tauri disks, our study provides support that this diversity could be driven by the radial redistribution of material from the cold outer to the warm inner regions, where the majority of known exoplanets are thought to form. In particular, large disks could form close-in gas giants that have super-solar but likely below unity C/O ratios. \rev{However, comprehensive studies combining multiple JWST programs and focusing on representative samples of disks with similar stellar properties are needed to determine whether prominent C$_2$H$_2$ emission is more prevalent in pebble-rich disks and, more broadly, to fully test pebble drift models. The developments of forward models should also be a priority to go beyond slab models retrieval and robustly convert the observed molecular emission to elemental abundances.}



\begin{acknowledgements}
The authors would like to thank the anonymous referee for valuable and detailed comments on the manuscript. B.T. thanks M. Benisty for interesting discussions. This work is based on observations made with the NASA/ESA/CSA James Webb Space Telescope. The data were obtained from the Mikulski Archive for Space Telescopes at the Space Telescope Science Institute, which is operated by the Association of Universities for Research in Astronomy, Inc., under NASA contract NAS 5-03127 for JWST. 

These observations are associated with program 1282. The following National and International Funding Agencies funded and supported the MIRI development: 
NASA; ESA; Belgian Science Policy Office (BELSPO); Centre Nationale d’Etudes Spatiales (CNES); Danish National Space Centre; Deutsches Zentrum fur Luft und Raumfahrt (DLR); Enterprise Ireland; 
Ministerio De Econom\'ia y Competividad; 
Netherlands Research School for Astronomy (NOVA); 
Netherlands Organisation for Scientific Research (NWO); 
Science and Technology Facilities Council; Swiss Space Office; 
Swedish National Space Agency; and UK Space Agency.

B.T. acknowledges support from the Programme National PCMI of CNRS/INSU with INC/INP co-funded by CEA and CNES.

M.S., K.S. and T.H. acknowledge support from the European Research Council under the Horizon 2020 Framework Program via the ERC Advanced Grant Origins 83 24 28. 

M.T., E.v.D., M.V. acknowledges support from the ERC grant 101019751 MOLDISK.

D.~S. was funded by the Deutsche Forschungsgemeinschaft (DFG, German Research Foundation) – project number: 550639632. This work is based on observations carried out under project
numbers L19ME with the IRAM NOEMA Interferometer. IRAM is supported by INSU/CNRS (France), MPG (Germany), and IGN (Spain).

I.K., A.M.A., and E.v.D. acknowledge support from grant TOP-1 614.001.751 from the Dutch Research Council (NWO). 

I.K. and J.K. acknowledge funding from H2020-MSCA-ITN-2019, grant no. 860470 (CHAMELEON).

E.v.D. acknowledges support from the Danish National Research Foundation through the Center of Excellence ``InterCat'' (DNRF150).

A.C.G. acknowledges support from PRIN-MUR 2022 20228JPA3A “The path to star and planet formation in the JWST era (PATH)” funded by NextGeneration EU and by INAF-GoG 2022 “NIR-dark Accretion Outbursts in Massive Young stellar objects (NAOMY)” and Large Grant INAF 2022 “YSOs Outflows, Disks and Accretion: towards a global framework for the evolution of planet forming systems (YODA)”.


D.G. thanks the Belgian Federal Science Policy Office (BELSPO) for the provision of financial support in the framework of the PRODEX Programme of the European Space Agency (ESA).

G.P. gratefully acknowledges support from the Max Planck Society.

J.V. is funded from the Hungarian NKFIH OTKA project no. K-132406, and this work was also supported by the NKFIH NKKP grant ADVANCED 149943. Project no.149943 has been implemented with the support provided by the Ministry of Culture and Innovation of Hungary from the National Research, Development and Innovation Fund, financed under the NKKP ADVANCED funding scheme.

T.K. acknowledges support from STFC Grant ST/Y002415/1

\end{acknowledgements}

\bibliographystyle{aa} 
\bibliography{export-bibtex-disk.bib} 

\newpage



\begin{appendix}

\section{Complementary material on slab model fit}
\label{app:slab_fit}
\begin{figure*}
\centering
\includegraphics[width=1.0\textwidth]{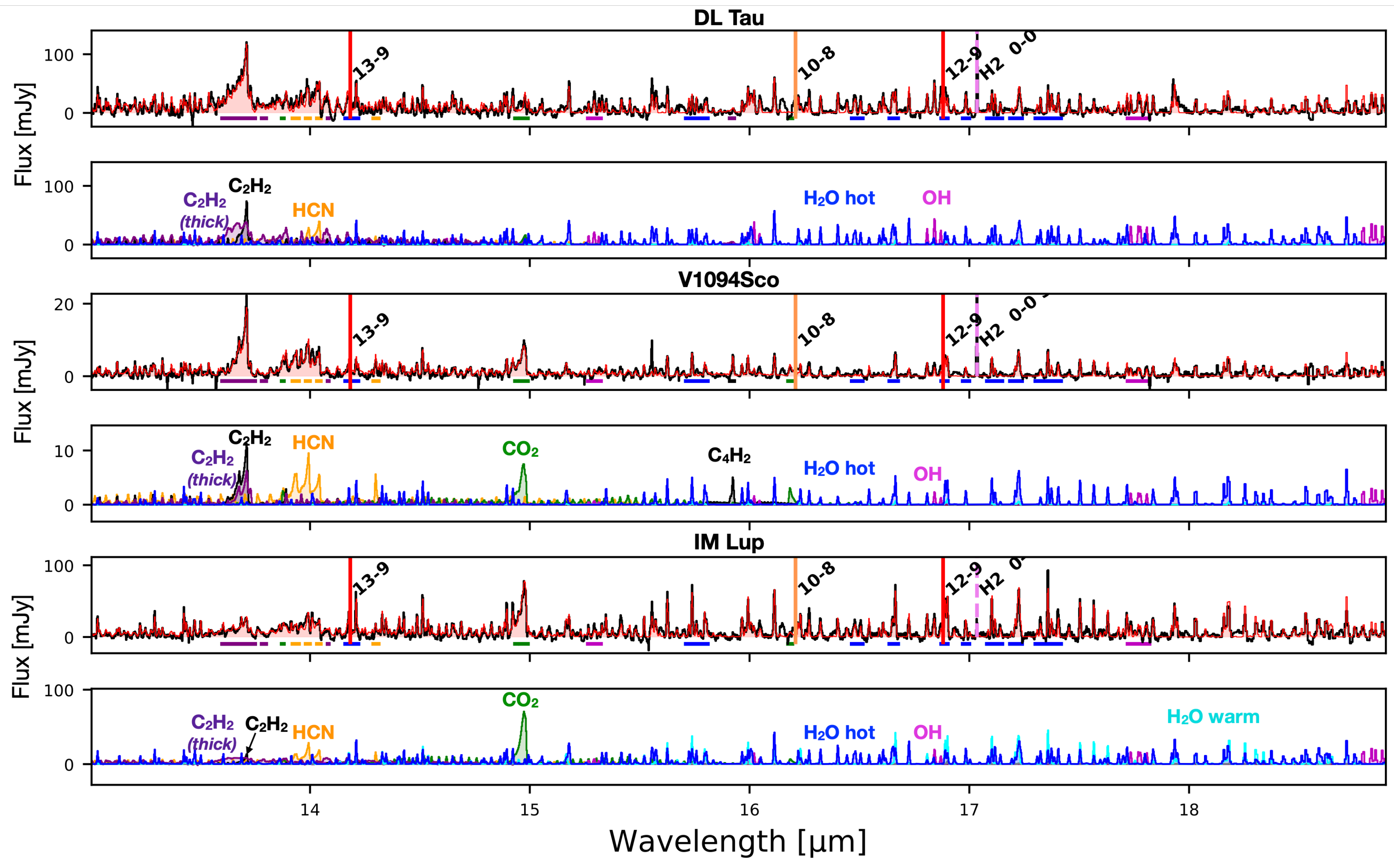}
\caption{Result of the slab model fit on the 12.5-19~$\mu$m region. For each source, the observed JWST spectrum is compared to the total model (red). The total emission model is further split into the different slab components. The best-fit slab models are reported in Table \ref{table:best_fit_slab}. The water slab models are in blue, cyan, and gray in decreasing excitation temperatures. The emission of C$_2$H$_2$ (optically thick component in purple and thinner in dark grey), OH (pink), CO$_2$ (green), and C$_4$H$_2$ (black, around 16$\mu$m for V1094 Sco) is also shown. The HI and H$_2$ lines are also labeled. The water emission in DL Tau and V1094 Sco is well fitted by two slabs with a dominance of hot water. The IM Lup water spectrum is matched by three slabs with excess cold water.}
\label{fig:appendix-water}
\end{figure*}

\rev{Tables \ref{table:best_fit_slab} and \ref{table:best_fit_slab2} provide the best fit parameters and their associated uncertainties obtained using the MCMC sampler \texttt{emcee}. We adopted a likelihood of \citet{2026MNRAS.545f2056K} with a multiplicative term to the observational uncertainties provided by the JWST pipeline to capture additional noise contributions. 15,000 steps have been used with flat priors and a burn-in phase of 2,000 steps. For HCN, CO$_2$, C$_4$H$_2$, and OH, one slab is used with 40 walkers. For C$_2$H$_2$ and H$_2$O two slabs are used with 42 walkers, except for IM Lup for which we fit the H$_2$O spectrum with three slabs and 60 walkers. The spectral windows used in the fit the emission of each species are shown in Fig. \ref{fig:appendix-water}. The fit of the C$_2$H$_2$ features by one versus two slabs is shown in Fig. \ref{fig:compa_C2H2-one-vs-two}.}

\begin{table}
\caption{Results of the slab model fits for C$_2$H$_2$ and H$_2$O.}
\label{table:best_fit_slab}
\centering
\small
\setlength{\tabcolsep}{3.5pt}
\begin{tabular}{lcccccc}
\toprule
 & \multicolumn{3}{c}{C$_2$H$_2$} & \multicolumn{3}{c}{H$_2$O} \\
\cmidrule(lr){2-4} \cmidrule(lr){5-7}
Source & $R$ [au] & $N$ [cm$^{-2}$] & $T$ [K] 
       & $R$ [au] & $N$ [cm$^{-2}$] & $T$ [K] \\
\midrule
DL Tau 
 & 0.08 & $19^{+0.8}_{-0.8}$ & $500^{+150}_{-150}$ 
 & 0.15 & $18.2^{+1}_{-0.5}$ & $900^{+200}_{-150}$ \\
 & 0.69 & $17.2^{+0.4}_{-0.2}$ & $200^{+50}_{-50}$ 
 & 0.52 & $18.2^{+0.5}_{-1}$ & $300^{+150}_{-150}$ \\
\addlinespace
IM Lup 
 & 0.04 & $19.6^{+0.75}_{-1.5}$ & $500^{+150}_{-150}$ 
 & 0.13 & $18.6^{+0.5}_{-1.5}$ & $800^{+200}_{-150}$ \\
 & 0.29 & $17.2^{+1}_{-0.25}$ & $200^{+150}_{-200}$ 
 & 0.42 & $18.6^{+1}_{-0.5}$ & $400^{+150}_{-150}$ \\
 &      &                      &                      
 & \textit{2.00} & \textit{17.8} & $200^{+150}_{-100}$ \\
\addlinespace
V1094Sco 
 & 0.15 & $18.4^{+1.5}_{-1}$ & $200^{+100}_{-100}$ 
 & 0.09 & $18.2^{+1}_{-0.5}$ & $600^{+200}_{-100}$ \\
 & \textit{0.13} & \textit{16} & $550^{+200}_{-200}$ 
 & 0.24 & $18.2^{+0.75}_{-1.5}$ & $300^{+100}_{-100}$ \\
\bottomrule
\end{tabular}
\end{table}

\begin{table*}
\caption{Results of the slab model fits for HCN, CO$_2$, and OH.}
\label{table:best_fit_slab2}
\centering
\small
\setlength{\tabcolsep}{4pt}
\begin{tabular}{lccccccccc}
\toprule
 & \multicolumn{3}{c}{HCN} & \multicolumn{3}{c}{CO$_2$} & \multicolumn{3}{c}{OH} \\
\cmidrule(lr){2-4} \cmidrule(lr){5-7} \cmidrule(lr){8-10}
Source 
& $R$ [au] & $\log N$ [cm$^{-2}$] & $T$ [K]
& $R$ [au] & $\log N$ [cm$^{-2}$] & $T$ [K]
& $R$ [au] & $\log N$ [cm$^{-2}$] & $T$ [K] \\
\midrule
DL Tau    
& $0.20^{+0.06}_{-0.03}$ & $17.2^{+0.4}_{-0.5}$ & $445^{+60}_{-90}$ 
& $0.69^{(a)}$ & $<15.7$ & $300^{(a)}$ 
& \textit{0.59} & \textit{15} & $1950^{+50}_{-40}$ \\

IM Lup    
& $0.09^{+0.03}_{-0.01}$ & $17.35^{+0.3}_{-0.4}$ & $655^{+65}_{-75}$ 
& $0.23^{+0.12}_{-0.07}$ & $16.8^{+0.4}_{-0.5}$ & $510^{+30}_{-40}$ 
& \textit{0.64} & \textit{15} & $1475^{+80}_{-80}$ \\

V1094 Sco 
& $0.04^{+0.01}_{-0.01}$ & $17.7^{+0.2}_{-0.2}$ & $680^{+60}_{-70}$ 
& $0.07^{+0.01}_{-0.01}$ & $18.8^{+0.2}_{-0.2}$ & $300^{+40}_{-30}$ 
& \textit{0.30} & \textit{15} & $1275^{+80}_{-100}$ \\
\bottomrule
\end{tabular}

\tablefoot{
(a) Parameters fixed to the warm water component to derive an upper limit on $N$(CO$_2$) for DL Tau.\\
Italic values correspond to optically thin emission where column density and emitting area are degenerate; uncertainties are therefore not provided.
}
\end{table*}


\begin{figure*}
\centering
\includegraphics[width=1.0\textwidth]{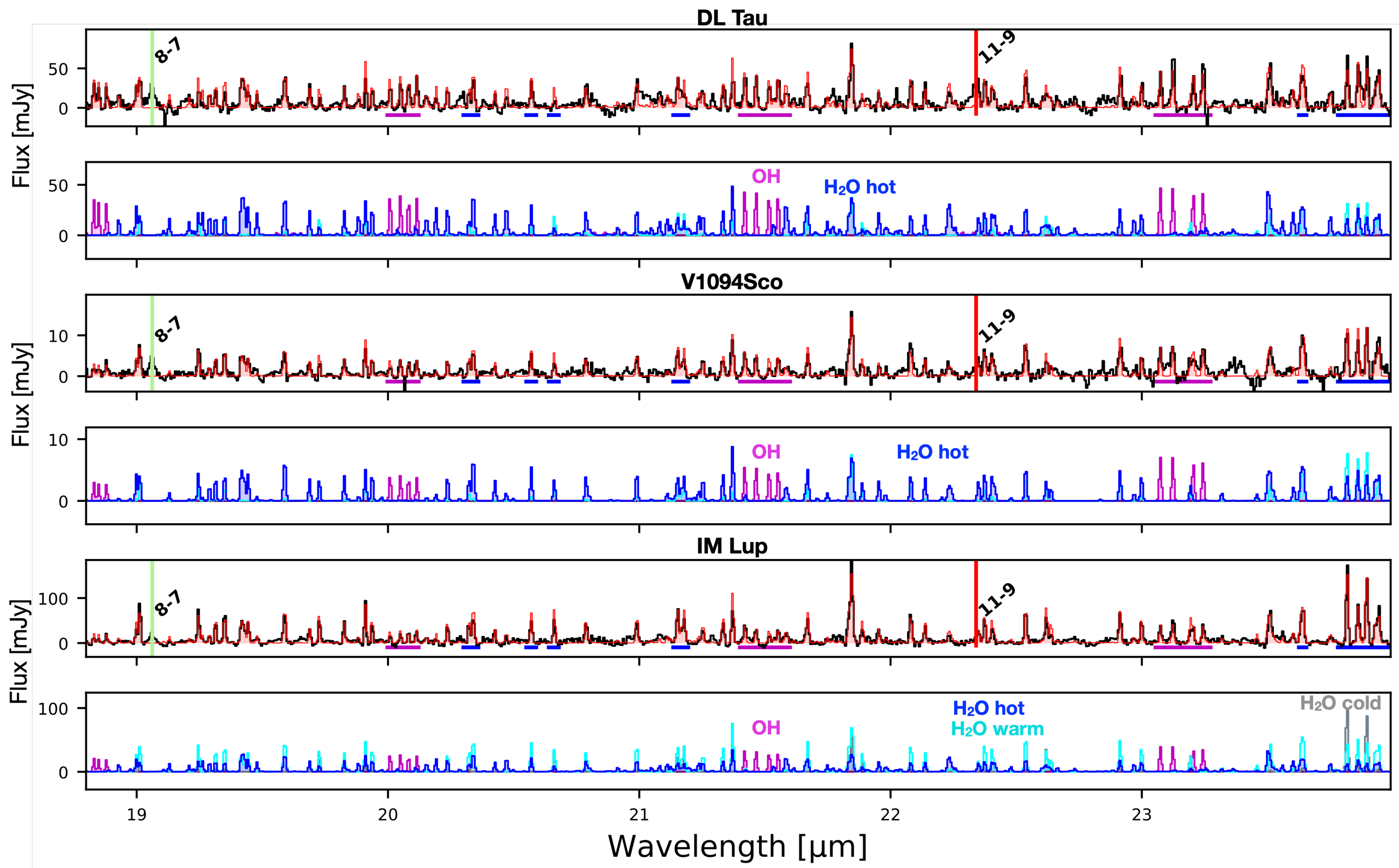}
\caption{Same as Figure \ref{fig:appendix-water2} but for the 19-24 $\mu$m range.}
\label{fig:appendix-water2}
\end{figure*}

\begin{figure}
\centering
\includegraphics[width=0.5\textwidth]{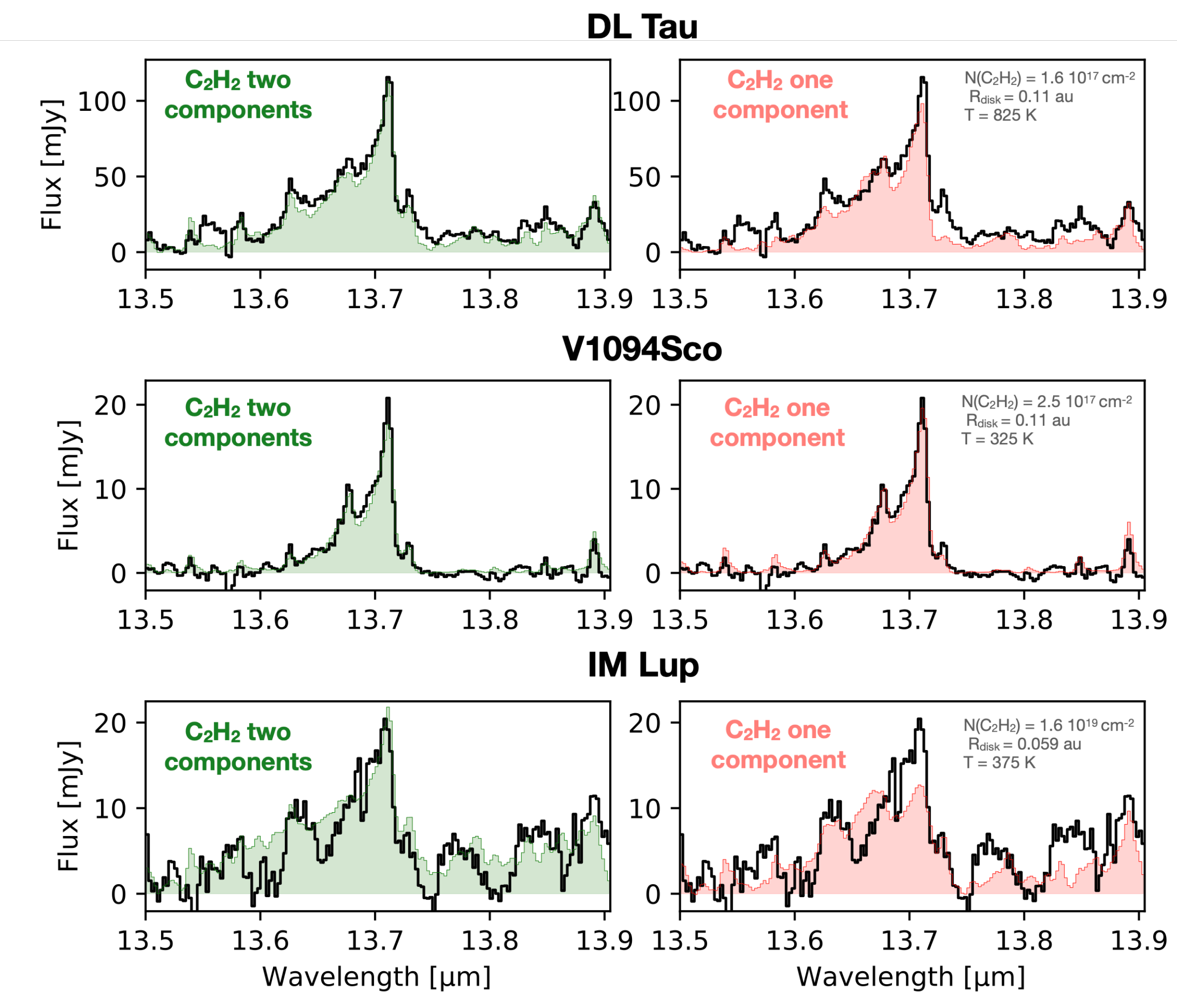}
\caption{\rev{Results of the fit of the C$_2$H$_2$ feature by two versus one slab. The observed spectra in black of DL Tau (top), V1094 Sco (middle), and IM Lup (bottom) have the contribution of H$_2$O, HCN, and CO$_2$ removed. Note that for V1094 Sco, a single slab is able to match most of the features, but the $Q$-branch of $^{13}$CCH$_2$. We recall that the slab model of C$_2$H$_2$ includes the contribution of the $^{13}$CCH$_2$ (see main text).}}
\label{fig:compa_C2H2-one-vs-two}
\end{figure}


\section{On the importance of underlying dust emission on the retrieved slab model parameters}

\label{appendix:RT}

It is often mentioned in the literature that the column density extracted from slab model fits corresponds to the column density of gas from the disk atmosphere down to the dust photosphere (i.e., where the dust emission becomes optically thick). In that case, dust extinction would be the limiting radiative transfer effect preventing us from probing the gas closer to the mid-plane. However, dust emission can also reduce and even cancel line emission, even if the emitting region of the molecular feature is much above and spatially distinct from the dust photosphere. This effect, discussed in the context of millimeter observations by \citet{2021MNRAS.501.3427R} and \citet{2022A&A...663A..58N}, is called continuum over-subtraction by the latter. To demonstrate the effect of continuum over-subtraction, we follow \citet{2022A&A...663A..58N} and approximate the expected vertical temperature gradient of a disk by two slabs: one of gas with an optical depth of $\tau_{gas}$ and one of optically thick dust located beneath it. The radiative transfer equation is then:
\begin{equation}
    I({\nu}) = \left( 1- e^{-\tau_{\text{gas}}(\nu)} \right) B_{\nu}(T_{\text{gas}}) + e^{-\tau_{\text{gas}}(\nu)}B_{\nu}(T_{\text{dust}}),
\end{equation}
where the first term accounts for the gas emission and the second term accounts for the dust emission absorbed by the slab of gas.
The continuum subtracted line intensity is then 
\begin{equation}
    I({\nu})-I_{\text{dust}}(\nu) = \left( 1- e^{-\tau_{\text{gas}}(\nu)} \right) \left( B_{\nu}(T_{\text{gas}}) - B_{\nu}(T_{\text{dust}}) \right).
\end{equation}
Compared with slab models without dust emission, the effect of continuum subtracted emission is therefore to reduce the line emission by a factor of:
\begin{equation}
f_{\text{over-sub}}(T_{\text{dust}},T_{\text{gas}},\lambda) = 1-B_{\nu}(T_{\text{dust}})/B_{\nu}(T_{\text{gas}}).
\end{equation}
We recover the textbook result that if the dust is hotter than the gas, the lines are seen in absorption. However, even if the dust is cooler than the gas, the emission of the molecular features is reduced as exemplified in Fig. \ref{fig:foversub}.

\section{A toy model for radial transport in halted pebble drift}

 \label{appendix:radial_transport}

\begin{figure*}
\centering
\includegraphics[width=0.85\textwidth]{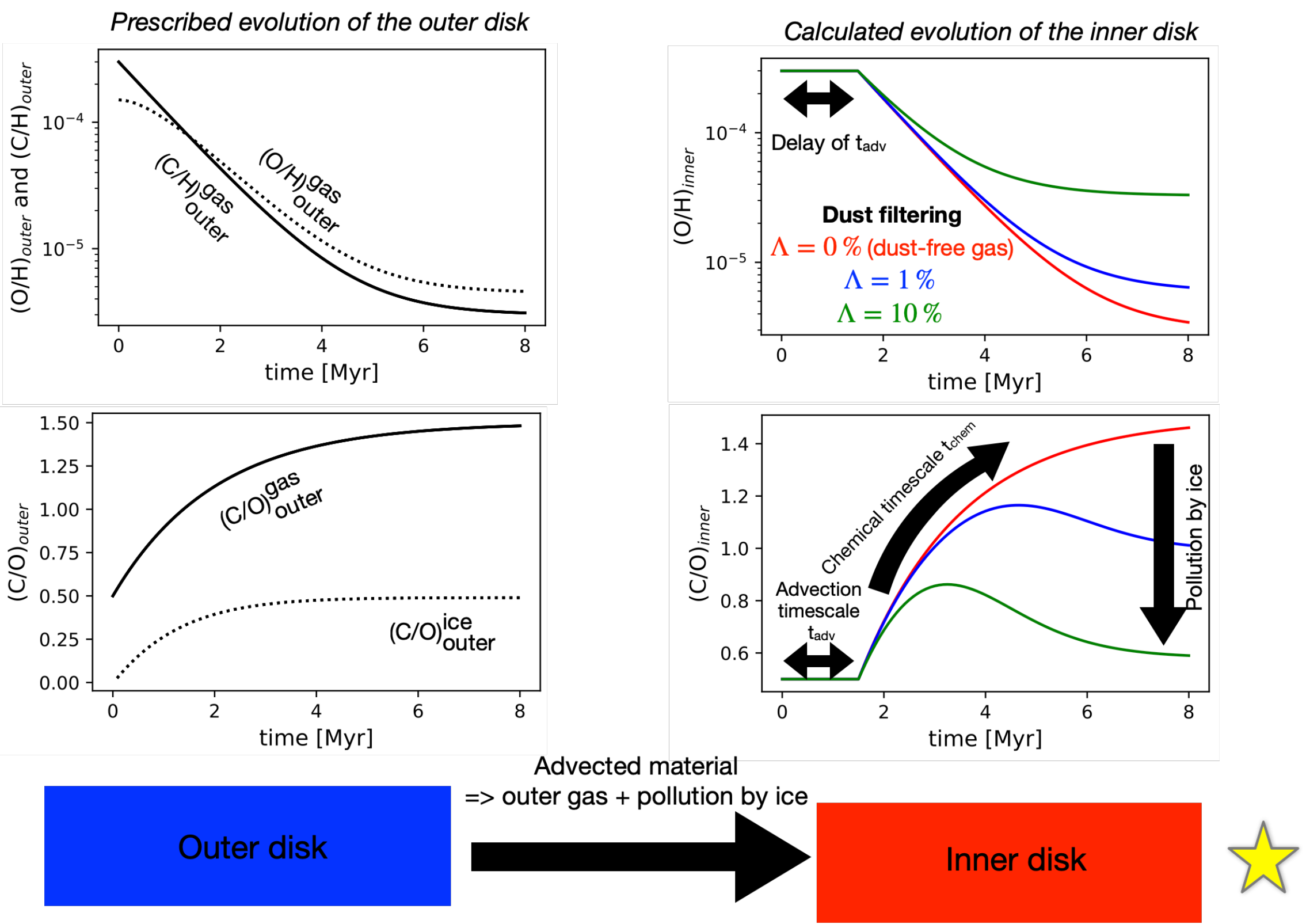}
\caption{Summary of the toy model developed in this work. We model the transport from outer to inner disk by two reservoirs. (Left) In the outer disk, the total C and O is constant but the partitioning between gas and ice varies with time as prescribed by  the variation of the C/O and O/H (black solid lines). (Right) The evolution of the inner disk is calculated by considering the composition of the gas in the outer disk and the pollution by icy grains parameterized by $\Lambda$, the gas-to-dust mass ratio of the advected material normalized to ISM value. The advection of the gs is mimicked by imposing a delay of $t_{\text{adv}}$ between the composition of the outer disk used to compute the composition of the advected material and that of the inner disk.}
\label{fig:appendix-toy-model}
\end{figure*}

\begin{figure}
\centering
\includegraphics[width=0.45\textwidth]{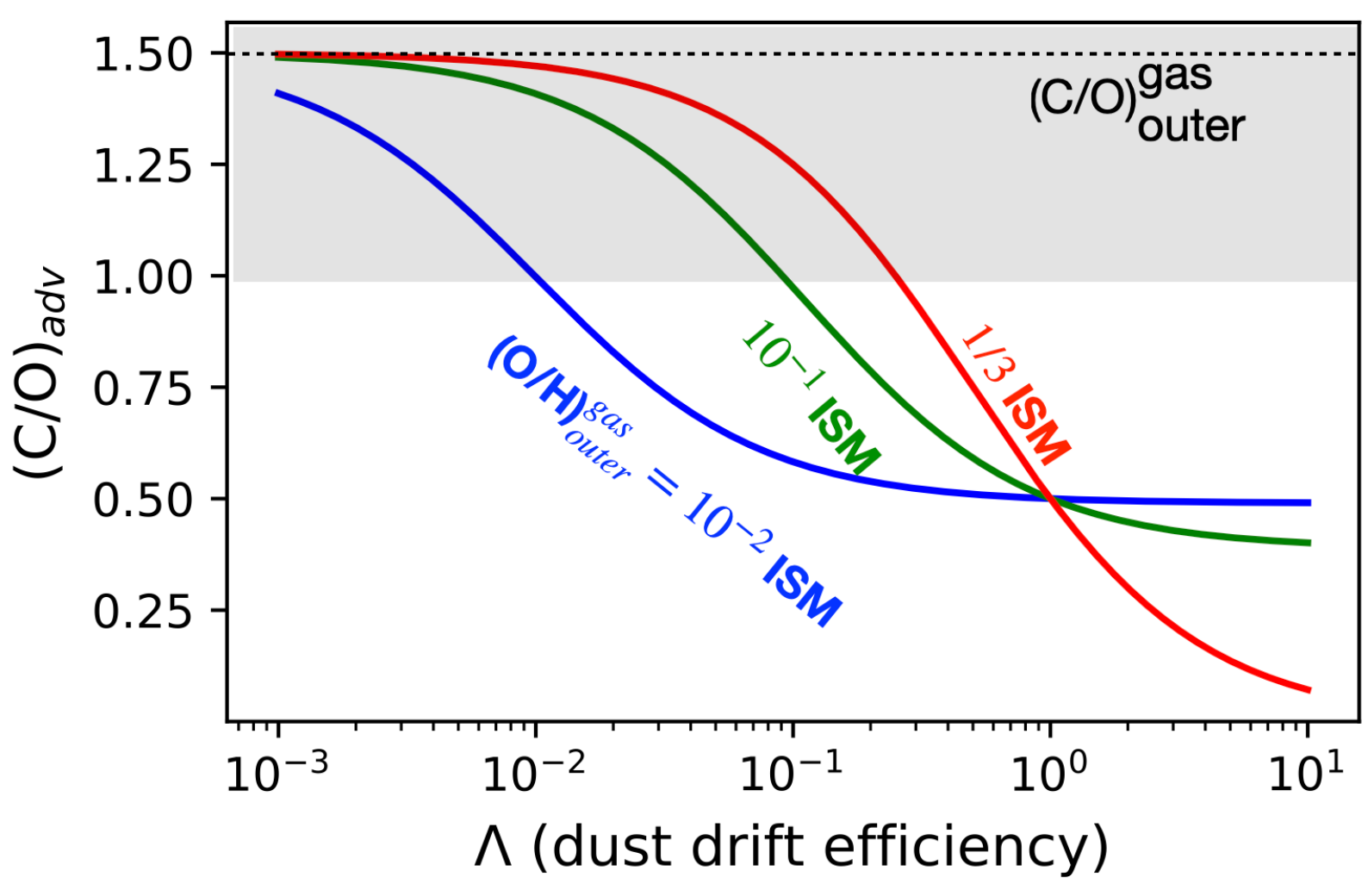}
\caption{Elemental composition of the gas plus ice advected to the inner disk depending on the efficiency to block the flow of icy dust and on the O/H ratio of the gas in the outer disk. For $\Lambda =1$, the material brought to the inner disk has an ISM gas-to-dust ratio and therefore has a solar (C/O) ratio, regardless of the partitioning of the elements between gas and dust. For low $\Lambda$, corresponding to limited pollution by icy grains, the composition approaches that of the gas in the outer disk. The effect of the pollution by icy grains is more prounced for low (O/H)$_{\text{outer}}^{gas}$ as ice carries more oxygen and carbon compared to the gas.}
\label{fig:appendix-toy-model-COadv}
\end{figure}

1-D models which consider the advection of gas and pebbles and include basic adsorption and desorption physics along with chemistry have been intensively developed over the past years \citep{2017MNRAS.469.3994B,2021ApJ...921...84K,2018A&A...611A..80B}. However, they tend to neglect the CO chemical conversion and trapping of volatiles in ices which leads to outer disks with elevated C/O and low (C+O)/H ratio in the gas. Here we design a toy model which mimics the transport of both the gas and dust and includes the observational constraints on outer disk chemistry. This toy model captures the essential features of the recent model of \citet{2025arXiv250711631S} in the case where the majority of the icy-grains are blocked in the outer disk. \rev{Five free parameters control the model: the chemical timescale, the advection timescale, the leakiness of the pebble traps, and the initial elemental abundances of carbon and oxygen. Here, we explore a limited number of parameters to highlight the essential predictions of the model.}

A summary of the model is presented in Fig. \ref{fig:appendix-toy-model}. We assume two reservoirs: the outer disk, where gas and volatile ice are present, and the inner disk, where all the volatiles are released in the gas phase. In the outer disk, we assume total volatile (gas and ice) elemental ratios of C/O=0.5, O/H$\equiv \mu^O = 3 \times 10^{-4}$, and C/H$\equiv \mu^C = 1.5 \times 10^{-4}$. We denote by (C/O)$_{\text{outer}}^{\text{gas}}$(t) and (O/H)$_{\text{outer}}^{\text{gas}}$(t) the carbon-to-oxygen and oxygen-to-hydrogen ratios of the gas in the outer disk which varies due to chemical reprocessing of CO and trapping of volatiles in the ice.  By conservation of elements, this allows us to calculate the carbon-to-oxygen ratio in the ice as:
\begin{equation}
\begin{split}
    \text{(C/O)}_{\text{outer}}^{\text{ice}} & =  \frac{\mu^{C} - \text{(C/H)}_{\text{outer}}^{\text{gas}} }  {\mu^{O} - \text{(O/H)}_{\text{outer}}^{\text{gas}}} = \frac{\mu^{C} - \text{(C/O)}_{\text{outer}}^{\text{gas}} \text{(O/H)}_{\text{outer}}^{\text{gas}} }  {\mu^{O} - \text{(O/H)}_{\text{outer}}^{\text{gas}}}.
    \end{split}
    \label{eq:appendix-C/O_ice}
\end{equation}

Instead of relying on a complex chemical model, we prescribe the time evolution of (C/O)$_{\text{outer}}^{\text{gas}}$(t) and (O/H)$_{\text{outer}}^{\text{gas}}$(t) following observational constraints gathered by \textit{Herschel} and ground-based millimeter observations. Specifically, we assume that these quantities transit from initial to final values on a chemical timescale of $t_{chem}$ (see Fig. \ref{fig:appendix-toy-model} top left):
\begin{equation}
\begin{split}
    \text{(O/H)}_{\text{outer}}^{\text{gas}}(t) &= \mu^O + \left( \text{(O/H)}_{\text{outer},\infty}^{\text{gas}}-  \mu^O \right) \left( 1 - e^{-t/t_{chem}} \right) \\
     \text{(C/O)}_{\text{outer}}^{\text{gas}}(t) &=  \mu^C/\mu^O + \left( \text{(C/O)}_{\text{outer},\infty}^{\text{gas}}-  \mu^C/\mu^O \right) \left( 1 - e^{-t/ 2t_{chem}} \right),
\end{split}
\label{eq:toy-model-1}
\end{equation}
where $ \text{(C/H)}_{\text{outer},\infty}^{\text{gas}}$ and $\text{(C/O)}_{\text{outer},\infty}^{\text{gas}}$ are the abundances ratios reached after few $t_{chem}$. In Fig. \ref{fig:toy-model-outer-inner} and \ref{fig:appendix-toy-model} we adopt a chemical timescale of $t_{\text{chem}}  = 1$~Myr, in line with observational estimates of the CO abundance in disks \citep{2020ApJ...891L..17Z} and chemical models \citep{2022ApJ...938...29F}. We note that $t_{\text{chem}}$ is only an effective chemical timescale tided to the adopted anzat of Eq (\ref{eq:toy-model-1}) and captures the result of several processes (sublimation and desorption, vertical transport, CO chemical conversion). The (O/H) converges to $\text{(O/H)}_{\text{outer},\infty}^{\text{gas}} = 3 \times 10^{-6}$, in line with the weak H$_2$O and CO lines measured in T Tauri disks \citep[e.g.,][]{2017ApJ...842...98D,2017A&A...599A.113M}. A final C/O of $\text{(C/O)}_{\text{outer},\infty}^{\text{gas}} = 1.5$ is adopted, typical of the values infered from the observations of small hydrocarbons in disks \citep{2016ApJ...831..101B,2019A&A...631A..69M}. Following the conservation of chemical elements, the (C/O) of the ice is given by Eq. (\ref{eq:appendix-C/O_ice}) (see Fig. \ref{fig:appendix-toy-model} bottom left).

We then consider that gas and icy-dust can be advected with different advection rates toward the inner disk. This amounts to take as a free parameter the gas-to-dust ratio of the material reaching the inner disk. We therefore denote by $\Lambda$ the gas-to-dust ratio of the advected material normalised by 100. The O and C abundances of the advected material are then
\begin{equation}
\begin{split}
    \text{(O/H)}_{\text{adv}} &=   \text{(O/H)}_{\text{outer}}^{\text{gas}}  + \left(  \mu^O - \text{(O/H)}_{\text{outer}}^{\text{gas}} \right) \Lambda \\ 
     \text{(C/H)}_{\text{adv}} &=   \text{(C/H)}_{\text{outer}}^{\text{gas}}  + \left(  \mu^C - \text{(C/H)}_{\text{outer}}^{\text{gas}} \right) \Lambda,
\end{split}
\label{eq:partitionning-OH-CH}
\end{equation}
which gives a C/O ratio of \rev{of the advected material of}
\begin{equation}
\begin{split}
    \rev{\text{(C/O)}_{\text{adv}} = \text{(C/O)}_{\text{outer}}^{\text{gas}}  \frac{1 + \left(  \mu^C/\text{(C/H)}_{\text{outer}^{\text{gas}}}  - 1 \right) \Lambda}{ 1+ (\mu^O/\text{(O/H)}_{\text{outer}}^{\text{gas}} -1 ) \Lambda } .}
\end{split}
\label{eq:partitionning-CO}
\end{equation}
We plot in Fig. \ref{fig:appendix-toy-model-COadv} the $\text{(C/O)}_{\text{adv}}$ ratio of the advected material for various amount of icy grains assuming a gas with high C/O and various values of O/H. For $\Lambda=1$, the advected material has a solar C/O abundances regardless of their partitioning between ice and gas  because gas and icy-grain are advected at the same pace. For low $\Lambda$ values, the advection of ice is negligible and the $\text{(C/O)}_{\text{adv}}$ is set by the high C/O of the gas. However, for low $\text{(O/H)}_{\text{outer}}^{\text{gas}}$, $\Lambda$ needs so be very small to reach the composition of the gas in the outer disk. This is because for low $\text{(O/H)}_{\text{outer}}^{\text{gas}}$, most of the oxygen and carbon are locked in ices. Therefore, a small fraction of ice in the advected material has a significant impact on the $\text{(C/O)}_{\text{adv}}$. This a necessary condition to obtain high (C/O) in the inner disk: for $\text{(O/H)}_{\text{outer}}^{\text{gas}} \simeq 10^{-2}$, in line with CO depletion factors inferred in T Tauri disks, and $\text{(C/O)}_{\text{outer}}^{\text{gas}} \simeq 1.5$, 99\% of the icy grains need to be retained in the outer disk to obtain a C/O ratio in the inner disk above 1; a condition that is likely never reached even in DL Tau and V1094 Sco (see discussion in Sec. \ref{subsec:elemental_abundance}). 


The other necessary condition to obtain elevated C/O in the inner disk is related the advection timescale, which is set by accretion processes.
In our toy model we take this process into account by introducing a delay between the composition of the outer disk used to compute the composition of the advected material and that of the inner disk. Therefore, Eq. (\ref{eq:partitionning-OH-CH}) and (\ref{eq:partitionning-CO}) gives for $t>t_{adv}$
\begin{equation}
\begin{split}
    \text{(O/H)}_{\text{inner}}^{\text{gas}}(t) &=   \text{(O/H)}_{\text{outer}}^{\text{gas}}(t-t_{\text{adv}})  + \left(  \mu^O - \text{(O/H)}_{\text{outer}}^{\text{gas}}(t-t_{\text{adv}})  \right) \Lambda \\ 
     \text{(C/O)}_{\text{inner}}^{\text{gas}}(t) &=   \text{(C/O)}_{\text{outer}}^{\text{gas}}(t-t_{\text{adv}})   + \left(  \mu^C - \text{(C/O)}_{\text{outer}}^{\text{gas}}(t-t_{\text{adv}})  \right) \Lambda,
\end{split}
\label{eq:toy-model-final-eq}
\end{equation}
where $\text{(O/H)}_{\text{outer}}^{\text{gas}}(t)$ and $\text{(C/O)}_{\text{outer}}^{\text{gas}}(t)$ are provided in Eq. (\ref{eq:toy-model-1}).

\begin{figure}
\centering
\includegraphics[width=0.5\textwidth]{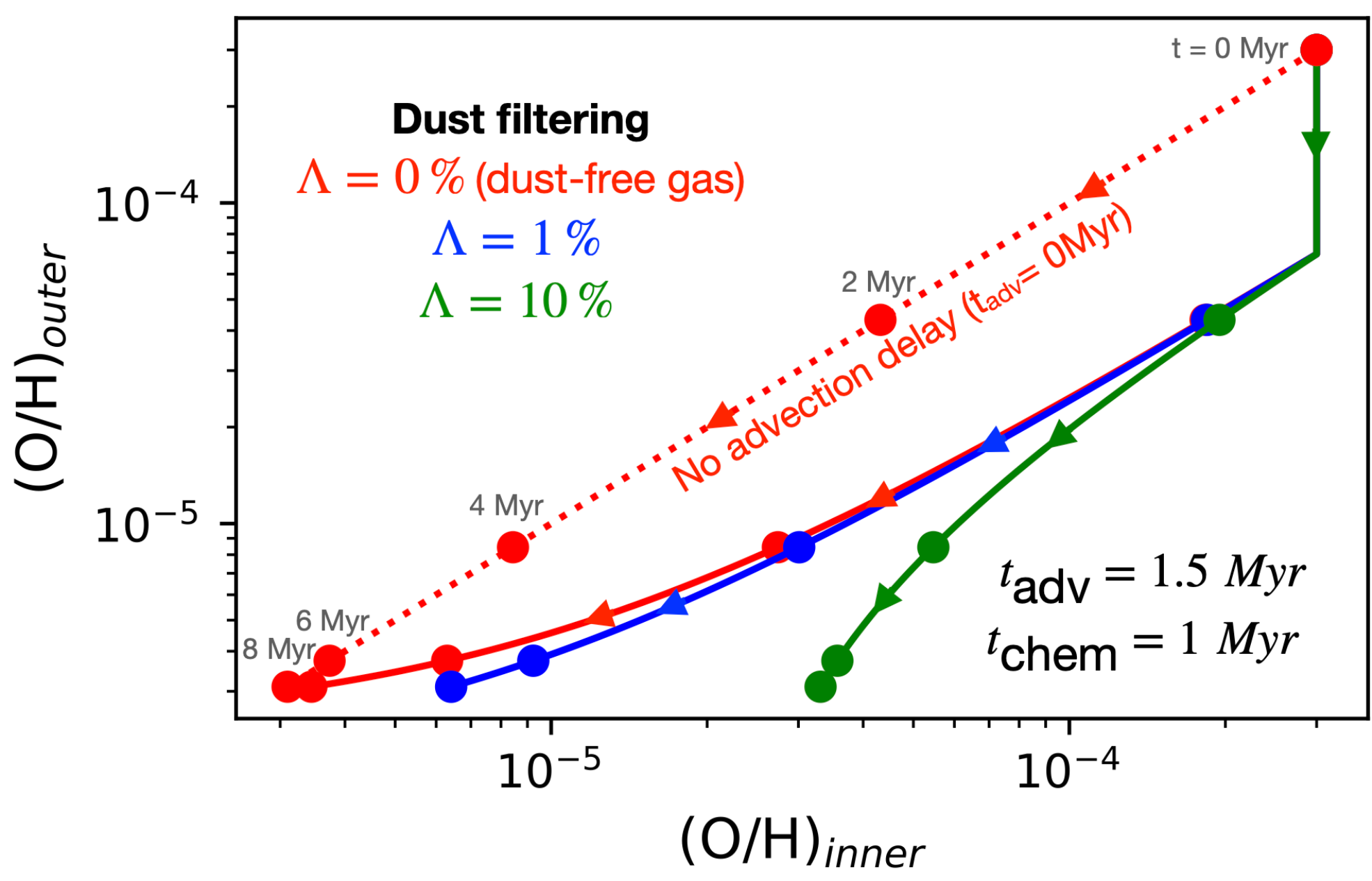}
\caption{Oxygen abundance of the outer versus inner disk predicted by our toy model of halted pebble drift. The O/H ratio in the gas of the outer versus inner disk is indicated every 2~Myr starting from solar values. Various values of dust filtering are shown (see color code in the figure).}
\label{fig:appendix-toy-model-inner-vs-outer-O-H}
\end{figure}

Figure \ref{fig:appendix-toy-model} (right panels) shows the resulting evolution of the inner disk for various values of $\Lambda$. After an advection timescale, the C/O rises and the O/H drops as the material in the outer disk is reprocessed on a chemical timescale. Interestingly, the C/O reaches a maximum before dropping to small but yet super-solar values. This corresponds to an optimum between high (O/H) in the gas, which limits the contribution of ice in the C and O budget, and high C/O in the gas which favors high C/O in the inner disk. The minimum O/H and maximum C/O depends on the pollution by icy-grains paramaterized by $\Lambda$. 
Figure \ref{fig:inner-vs-outer} is a combination of the bottom two plots. In Fig. \ref{fig:appendix-toy-model-inner-vs-outer-O-H} we show a similar figure but for the (O/H) ratio. 

\section{C$_2$H and C$^{18}$O line fluxes}
\label{sec:appendix-C2H-C18O-flux-measurments}

In Sec. \ref{sec:discussion}, we use the C$_2$H(3-2)/C$^{18}$O(2-1) line fluxes collected from various work to discuss the connection between outer and inner disk chemistry. Table \ref{table:appendix-C2H} summarizes the references associated with each flux. The \ce{C2H}(3-2) transition is split into four hyperfine components and we take the flux summed over all these components. For Sz 98, \citet{2019A&A...631A..69M} cover only two of them: the (N = 3–2; J = 7/2-5/2; F = 4–3 and F = 3–2) transitions. Based on the typical flux ratio of the different hyperfine components measured by \citet{2019ApJ...876...25B}, the total flux of Sz 98 is estimated by multiplying by a factor 1.4 the flux of the (N = 3–2; J = 7/2-5/2; F = 4–3 and F = 3–2) transition.

\begin{table}[h]
\begin{center}
 \caption{Millimeter line fluxes used in Fig. \ref{fig:inner-vs-outer}.}
\begin{tabular}{ |c|c  c c c| } 
 \hline
     Source   &  C$^{18}$O(2-1) flux & Ref.  & \ce{C2H}(3-2)  flux & Ref.\\ 
       &  mJy km/s & & mJy km/s  & \\ 
     \hline
IM Lup    & 1,225 $\pm$ 120   &   (a) & 2,448 $\pm$ 214   &   (a)  \\
Sz 98     & $<$180            &   (e) & 1,540 $\pm$  51   &   (b)  \\
CY Tau    &  220 $\pm$ 10     &   (c) & 1,850 $\pm$ 120   &   (d)  \\
DL Tau    &  80  $\pm$ 10     &   (c) & 780   $\pm$ 50    &   (d)  \\
DN Tau    &  90  $\pm$ 10     &   (c) & 1,380 $\pm$ 40    &   (d)  \\
IQ Tau    &  70  $\pm$ 10     &   (c) & 2,300 $\pm$ 110   &   (d)  \\
CI Tau    &  840 $\pm$ 30     &   (c) & 1,890 $\pm$ 130   &   (d)  \\
DR Tau    &  622 $\pm$ 10     &   (f) &   24  $\pm$ 5     &   (g)  \\
 \hline
\end{tabular}
\label{table:appendix-C2H}
\end{center}
{(a) \citet{2019A&A...631A..69M}   (b) \citet{2019ApJ...876...25B} (c) \citet{2024A&A...685A.126S} (d) Semenov et al. in prep.  (e) \citet{2016ApJ...828...46A}  (f) \citet{2023ApJ...943..107H} (g) \citet{2024ApJ...973..135H} }
\end{table}

\section{Extended H$_2$ emission}
\label{sec:appendix-H2-maps}
In this appendix, we provide the maps of the H$_2$ emission focusing on the detected lines. The emission of warm dust from the inner disk has been removed by subtracting a map computed from the adjacent line-free spectral maps as detailed in \citet{2025arXiv250802576K}. We reveal extended emission in the three sources with emission shifted perpendicular to the disk major axis. This morphology indicates an origin of the H$_2$ in an outflow and constitute promissing candidates for molecular gas directly launched from the disk surface. Less excited lines show more extended emission indicative of temperature and density gradients. In IM Lup, the outflow origin is less clear and H$_2$ emission might trace the disk surface. \rev{The effect of the obscuration by the large disks is particularly clear for V1094 Sco (see H$_2$ S(1) emission).}

\begin{figure}
\centering
\includegraphics[width=0.5\textwidth]{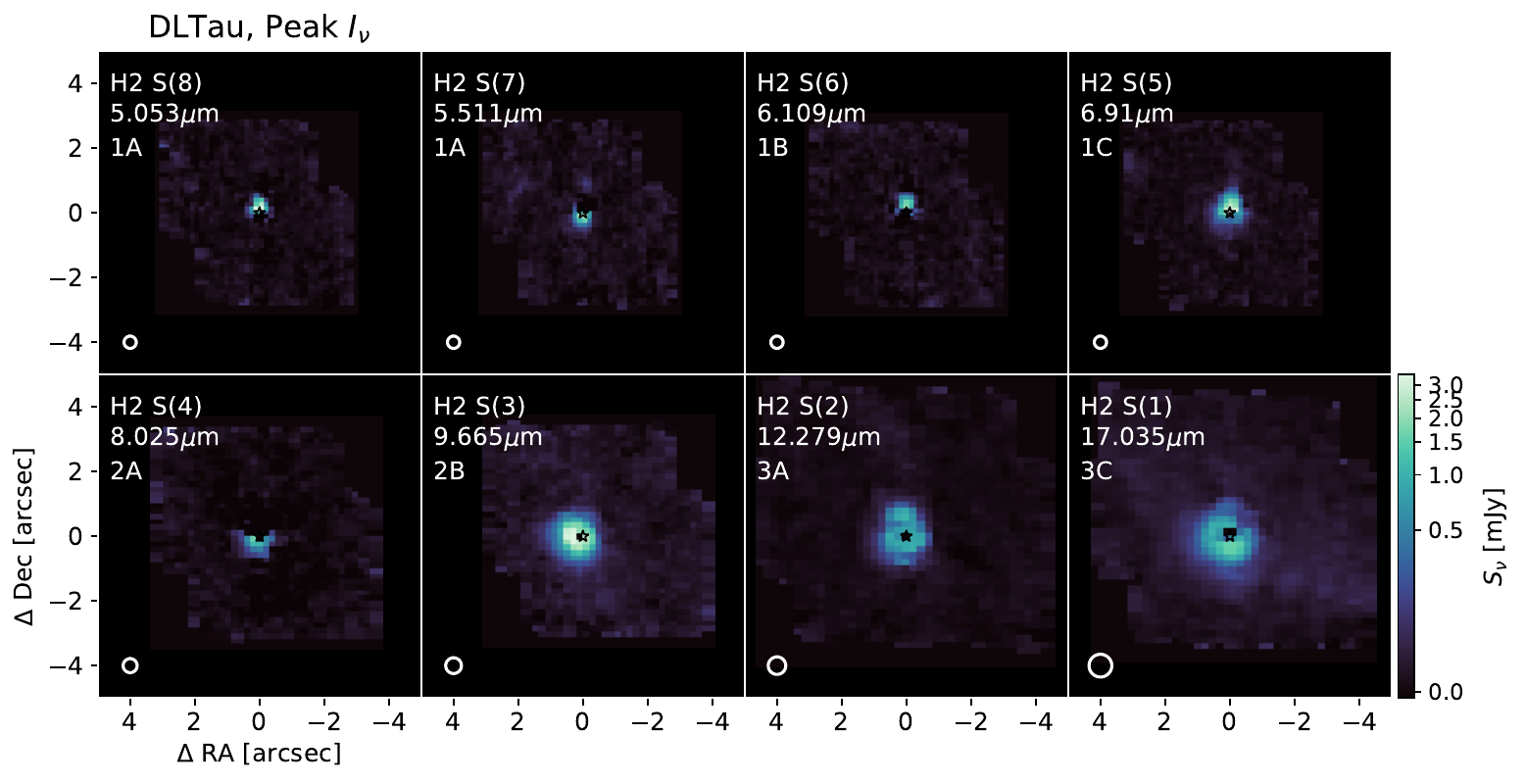}
\caption{H$_2$ intensity map of the rotational lines of DL Tau. The units of the maps are Jansky per pixel.}
\label{fig:H2_map}
\end{figure}

\begin{figure}
\centering
\includegraphics[width=0.5\textwidth]{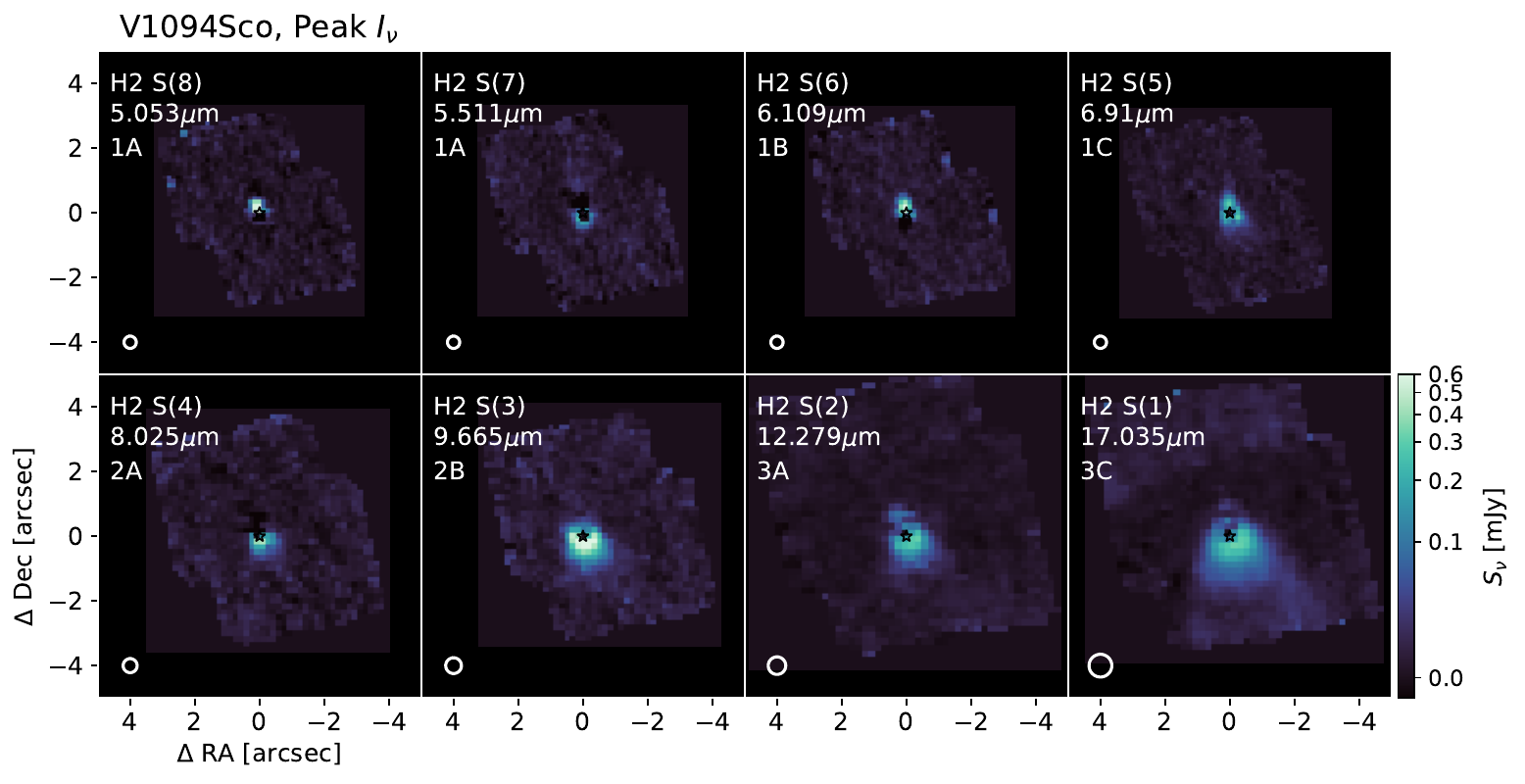}
\caption{H$_2$ intensity map of the rotational lines of V1094Sco.}
\label{fig:H2_map}
\end{figure}

\begin{figure}
\centering
\includegraphics[width=0.5\textwidth]{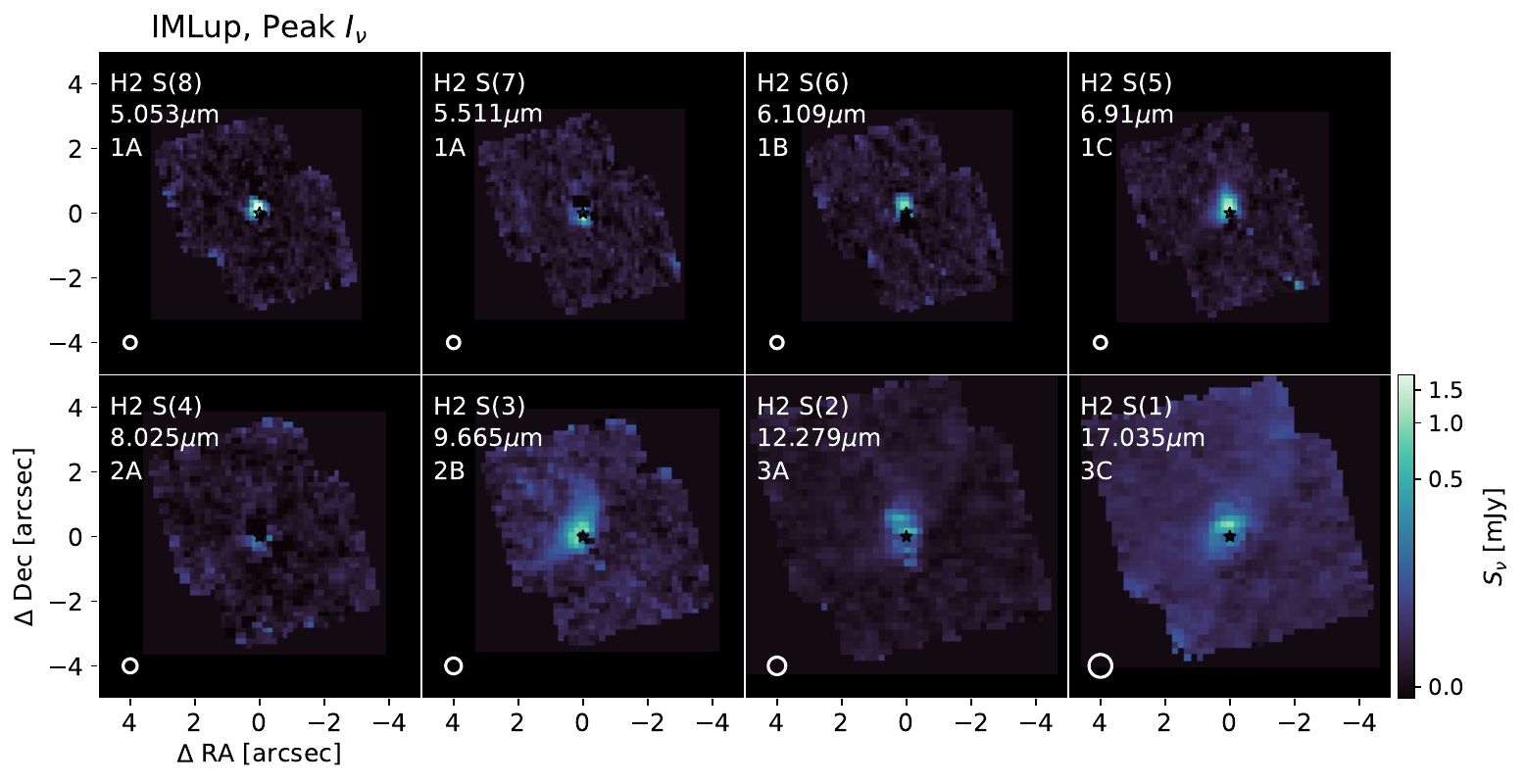}
\caption{H$_2$ intensity map of the rotational lines of IM Lup.}
\label{fig:H2_map}
\end{figure}

\end{appendix}

\end{document}